\documentclass[%
 aip,
 jmp,%
 amsmath,amssymb,
 reprint,%
]{revtex4-1}

\usepackage{graphicx}% Include figure files
\usepackage{dcolumn}% Align table columns on decimal point
\usepackage{bm}% bold math
\begin{document}

\preprint{AIP/123-QED}

\title[Fluid vs. kinetic magnetic reconnection with strong guide fields]{Fluid vs. kinetic magnetic reconnection with strong guide fields}

\author{A. Stanier}
\email{stanier@lanl.gov}
\author{Andrei N. Simakov}
\author{L. Chac\'on}
\author{W. Daughton}

\affiliation{Los Alamos National Laboratory, Los Alamos, New Mexico 87545, USA}

\date{\today}

\begin{abstract}The fast rates of magnetic reconnection found in both nature and experiments are important to understand theoretically. Recently, it was demonstrated that two-fluid magnetic reconnection remains fast in the strong guide field regime, regardless of the presence of fast-dispersive waves. This conclusion is in agreement with recent results from kinetic simulations, and is in contradiction to the findings in an earlier two-fluid study, where it was suggested that fast-dispersive waves are necessary for fast reconnection. In this paper, we give a more detailed derivation of the analytic model presented in a recent letter, and present additional simulation results to support the conclusions that the magnetic reconnection rate in this regime is independent of both collisional dissipation and system-size. In particular, we present a detailed comparison between fluid and kinetic simulations, finding good agreement in both the reconnection rate and overall length of the current layer. Finally, we revisit the earlier two-fluid study, which arrived at different conclusions, and suggest an alternative interpretation for the numerical results presented therein.  \end{abstract}

\pacs{Valid PACS appear here}% PACS, the Physics and Astronomy
%Use showkeys class option if keyword
\maketitle

\section{\label{sec:intro}Introduction}

Magnetic reconnection~\cite{priest00,zweibel09} is the changing of magnetic field-line connectivity by localised magnetic flux unfreezing. This process can release vast amounts of stored magnetic energy, resulting in rapid heating and particle acceleration within solar flares,~\cite{priest00,su13} coronal mass ejections, and substorms in the Earth's magnetosphere.~\cite{angelopoulos08}

Reconnection is a common process in magnetically confined laboratory plasmas, where it can be utilised to reach desirable equilibrium pressure and magnetic field profiles~\cite{baker84,ono11,stanier13,browning14} during plasma start-up. However, reconnection can also be undesirable in tokamaks and other devices, where it can destroy nested flux-surfaces and degrade confinement.~\cite{wesson90,yamada94,chapman10}

In these applications, the details of the reconnection vary, depending on the magnetic field geometry and collisionality of the plasma. Low-$\beta$ reconnection, where the plasma-$\beta$ is the ratio of thermal to magnetic pressures, occurs when the in-plane reconnecting magnetic field is dominated by a strong out-of-plane \textit{guide} magnetic field component. This reconnection regime is pertinent to tokamaks, the solar corona, and magnetically dominated astrophysical environments. However, despite the wide range and importance of these applications, reconnection in this regime remains poorly understood.

An outstanding theoretical question concerns the fast timescales of magnetic energy release, compared with collisional timescales. It has been demonstrated numerically~\cite{mandt94,ma96,biskamp97,birn01} and analytically~\cite{chacon07,simakov08,simakov09} that zero guide-field  reconnection can be fast; a significant fraction of the Alfv\'enic rate and independent of both dissipation~\cite{biskamp97,shay98,hesse99} and system-size,~\cite{shay99,huba04} provided that the dissipation region (DR) thickness $\delta$ falls below the ion-skin depth $d_i = c/\omega_{pi}$. Here, $c$ is the speed-of-light, $\omega_{pi}$ is the ion plasma frequency, and the DR is the region where collisional dissipation is the dominant contribution to the parallel electric field associated with reconnection. With a finite guide-field, it has been shown that the ion-sound Larmor radius $\rho_s = \sqrt{T_e/m_i}/\Omega_{ci}$, where $T_e$ is the electron temperature, $m_i$ the ion mass, and $\Omega_{ci}$ the ion cyclotron frequency, can play the role of the ion-skin depth as the threshold for fast reconnection.~\cite{aydemir92,kleva95,bhattacharjee05,schmidt09,rogers01,simakov10}  It is also worth noting that time-dependent studies with a guide-field have found faster than exponential tearing growth-rates in the non-linear regime when finite electron inertia,~\cite{ottaviani93} or finite ion gyro-radius,~\cite{grasso00,loureiro08,comisso13} effects are included.

If the two-fluid scales $d_i$ or $\rho_s$ are large enough, the plasma can support fast-dispersive waves (FDWs) with frequency $\omega \propto k^2$ for wavenumber $k$, namely whistler and kinetic Alfv\'en waves, respectively. It was proposed in Ref.~\onlinecite{rogers01} that these FDWs play a critical role in facilitating fast reconnection, where the outflow velocity from the DR scales as the phase-velocity of the wave, $v_{\textrm{o}}\propto k \propto 1/\delta$, to give dissipation-independent electron flux through the DR: $v_{\textrm{o}}\delta\propto \delta^0$.

In Fig.~3a of Ref.~\onlinecite{rogers01} it was further demonstrated that the reconnection rate decreases with the plasma-$\beta$ for fixed mass-ratio and ratio of guide-to-reconnecting field. For very low-$\beta$, such that $\beta/2 \leq m_e/m_i$ with $m_e$ the electron mass, $\rho_s$ falls below the electron skin-depth, $d_e = \sqrt{m_e/m_i}d_i$, and there are no FDWs supported in the plasma. Due to the decreased rate and opening angle between the magnetic separators, it was concluded that reconnection is slow in this regime. However, we note that no formal scaling study against dissipation or system-size was presented in Ref.~\onlinecite{rogers01}, and recent results have called this latter conclusion into question. Firstly, it has been demonstrated that reconnection is fast in electron-positron plasmas,~\cite{bessho05,daughton07,chacon08,liu15} which do not support FDWs. Secondly, kinetic simulations~\cite{liu14,tenbarge14} have shown that reconnection remains fast in low-$\beta$ ion-electron plasmas when $\beta/2 < m_e/m_i$, appearing to contradict earlier two-fluid results.~\cite{rogers01}

In this paper, we address this contradiction with analytical and numerical results, demonstrating that reconnection is indeed fast within a low-$\beta$ two-fluid model in the absence of FDWs. A brief summary of our main results in support of this conclusion has been presented in a recent letter.~\cite{stanier15} In this paper, we give a more complete description of the numerical and analytical methods, and present further results that support the conclusions of dissipation and system-size independence that we were unable to present in Ref.~\onlinecite{stanier15} due to space limitations. We also present new results showing detailed comparisons of the reconnection rate, magnetic flux, and current density, between fluid and fully kinetic simulations. Good agreement is found between fluid and kinetic simulations in cases with and without FDWs, supporting the conclusion that the reconnection rate is independent of the DR physics in both cases. Finally, we perform simulations that can be more easily compared with those presented in Ref.~\onlinecite{rogers01}, and suggest an alternative interpretation for the observed decrease in reconnection rate with plasma-$\beta$.

The layout of the paper is as follows. In Sec.~\ref{sec:tfmodel}, we describe the low-$\beta$ two-fluid model that forms the basis of the discrete analysis and numerical simulations in this paper. In Sec.~\ref{sec:discretemodel}, we discretise this model at the DR, and show analytically how the DR must self-adjust to permit fast, dissipation independent reconnection. The numerical method used to solve the model equations of Sec.~\ref{sec:tfmodel} is described in Sec.~\ref{sec:method}, then in Sec.~\ref{sec:numresults} we demonstrate that the DR does have the capacity to self-adjust in the manner required to give dissipation independent rates in cases with and without FDWs. In this section, we also compare and find excellent agreement between fluid and kinetic descriptions. It is further shown, in Sec.~\ref{sec:sys-size}, that rates are a significant fraction of the Alfv\'enic rate and independent of system-size, so that reconnection is formally fast both with and without FDWs. The comparison with Ref.~\onlinecite{rogers01} is given in Sec.~\ref{sec:rogerscomparison}, before conclusions are drawn in Sec.~\ref{sec:conclude}.

\section{\label{sec:tfmodel}Low-$\beta$ two-fluid model}
The low-$\beta$ ($\beta{\ll}1$), two-dimensional (${\partial_z{=}0}$, where ${\boldsymbol{\hat{z}}{=}\boldsymbol{\nabla}z}$ is the out-of-plane direction), two-field reconnection model~\cite{schep94,kleva95,cafaro98,bhattacharjee05,schmidt09,simakov10} can be derived from the two-fluid magnetised plasma equations~\cite{braginskii65} assuming uniform temperatures with cold-ions ($\boldsymbol{\nabla}T_i {=} \boldsymbol{\nabla}T_e {=} 0$ and $T_i {\ll} T_e$ for ion temperature $T_i$), strong-guide field ($|\boldsymbol{B}| {\ll} B_0$ where $\boldsymbol{B}{=}\boldsymbol{\hat{z}}{\times} \boldsymbol{\nabla}\psi$ is the in-plane field, $\psi$ is the flux, and $B_0 \boldsymbol{\hat{z}}$ is the guide-field), an MHD ordering (the electric drift velocity is on the order of the ion thermal speed $\boldsymbol{v}_E {\sim} v_{Ti}$) and small ion parallel flow ($v_{\parallel i} {\ll} v_{Ti}$). As is typically done (see e.g. Ref.~\onlinecite{biskamp00}), we also assume a simple perpendicular viscosity closure for both ions and electrons. Here, the equations are given in terms of the vector magnetic field $\boldsymbol{B}$ for the analysis that follows, and are normalised by a characteristic Alfv\'en velocity and macroscopic length-scale.
\begin{equation}\label{vorteqn}\left(\partial_t + \boldsymbol{v} \cdot \boldsymbol{\nabla}\right)\omega = \boldsymbol{B} \cdot \boldsymbol{\nabla} j + \mu \nabla^2\omega,\end{equation}
\begin{eqnarray}\label{vecBeqn}\partial_t \boldsymbol{B}^* - \boldsymbol{\nabla}\times\left(\boldsymbol{v}\times \boldsymbol{B}^*\right) &=& \rho_s^2 \boldsymbol{\nabla} \times \left[\boldsymbol{B} \times \left(\boldsymbol{\hat{z}}\times \boldsymbol{\nabla}\omega\right)\right] \\\nonumber &-& \boldsymbol{\nabla} \times \left[\boldsymbol{\nabla} \times \left(\eta \boldsymbol{B} - \eta_H\nabla^2 \boldsymbol{B}\right)\right], \end{eqnarray}
where $\boldsymbol{v} = \boldsymbol{\hat{z}}\times \boldsymbol{\nabla} \phi$ is the in plane velocity in terms of streamfunction $\phi$, $\omega = \nabla^2 \phi$ is the vorticity, $j = \nabla^2 \psi$ is the parallel current density and $\boldsymbol{B}^* = \boldsymbol{B}+d_e^2\boldsymbol{\nabla}\times\left(\boldsymbol{\nabla}\times \boldsymbol{B}\right)$. The normalised dissipation coefficients are the ion collisional viscosity $\mu$, the plasma resistivity $\eta$, and the hyper-resistivity $\eta_H$ that can be written in terms of an electron collisional viscosity $\mu_e$ as $\eta_H=d_i^2 \mu_e$.

When $d_e {=} \eta {=} \eta_H {=} 0$, Eq.~(\ref{vecBeqn}) can be written in flux form as $\partial_t\psi {+} \boldsymbol{v}_s{\cdot} \boldsymbol{\nabla} \psi {=} 0$, where $\boldsymbol{v}_s {=} \boldsymbol{\hat{z}} {\times} \boldsymbol{\nabla}\left(\phi{-}\rho_s^2\nabla^2\phi\right)$ is the velocity that advects the frozen-in magnetic flux.~\cite{schmidt09,grasso99} Physically, this is an electron perpendicular velocity that combines the lowest-order electric drift $\boldsymbol{v}$ (the single-fluid velocity in this model) and an electron diamagnetic drift that can become significant close to the X-point due to the characteristic density asymmetry pattern~\cite{kleva95,stanier13} associated with strong guide-field reconnection.

Relaxing the $d_e {=} 0$ condition gives $\partial_t\psi {+} \boldsymbol{v}_s{\cdot} \boldsymbol{\nabla} \psi  =  d_e^2\left(\partial_t j + \boldsymbol{v}\cdot \boldsymbol{\nabla}j\right)$. Here, in the last term on the right, we have $\boldsymbol{v}\cdot \boldsymbol{\nabla} j$ rather than $\boldsymbol{v}_s\cdot \boldsymbol{\nabla} j$, since the diamagnetic contribution to electron inertia cancels out with the lowest order electron gyro-viscosity (see Ref.~\onlinecite{hazeltine04}). Thus, we avoid the potentially large and spurious term that is present in non-reduced Hall-MHD simulations that include electron inertia while neglecting electron gyro-viscosity. 

In a uniform ($\boldsymbol{B} = B_0 \boldsymbol{\hat{z}}$) collisionless ($\eta{=}\mu{=}\eta_H{=}0$) plasma, Eqs.~(\ref{vorteqn}) and (\ref{vecBeqn}) support waves with dispersion relation 
\begin{equation}\label{dispersion}\omega = k_\parallel \sqrt{(1+\rho_s^2k^2)/(1+d_e^2k^2)},\end{equation}
where $k_\parallel {=} \boldsymbol{\hat{z}}{\cdot} \boldsymbol{k}$. In the limit $\rho_s^2 k^2\, {\gg}\, 1 \,{\gg}\, d_e^2 k^2$, Eq.~(\ref{dispersion}) describes the fast-dispersive kinetic Alfv\'en wave with $\omega {=} \rho_s k_\parallel k$. However, for $\rho_s \leq d_e$ there are no FDWs. There is only one condition here for the presence of FDWs, in place of the two-conditions discussed in Ref.~\onlinecite{rogers01}, since the condition for the presence of whistler waves has been ordered out due to the strong guide field assumption. The condition $\rho_s \leq d_e$ is equivalent to $\beta/2 \leq m_e/m_i$. 

\section{\label{sec:discretemodel}Discrete model}

A summary of this section can be found in the recently published letter.~\cite{stanier15} Here, we additionally include the intermediate steps in the derivation of the governing equation for the DR thickness, and highlight the similarities between strong guide field reconnection in ion-electron plasmas without FDWs, and reconnection in the low-$\beta$ pair-plasma regime studied previously in Ref.~\onlinecite{chacon08}.

\begin{figure}
\includegraphics[width=0.4\textwidth]{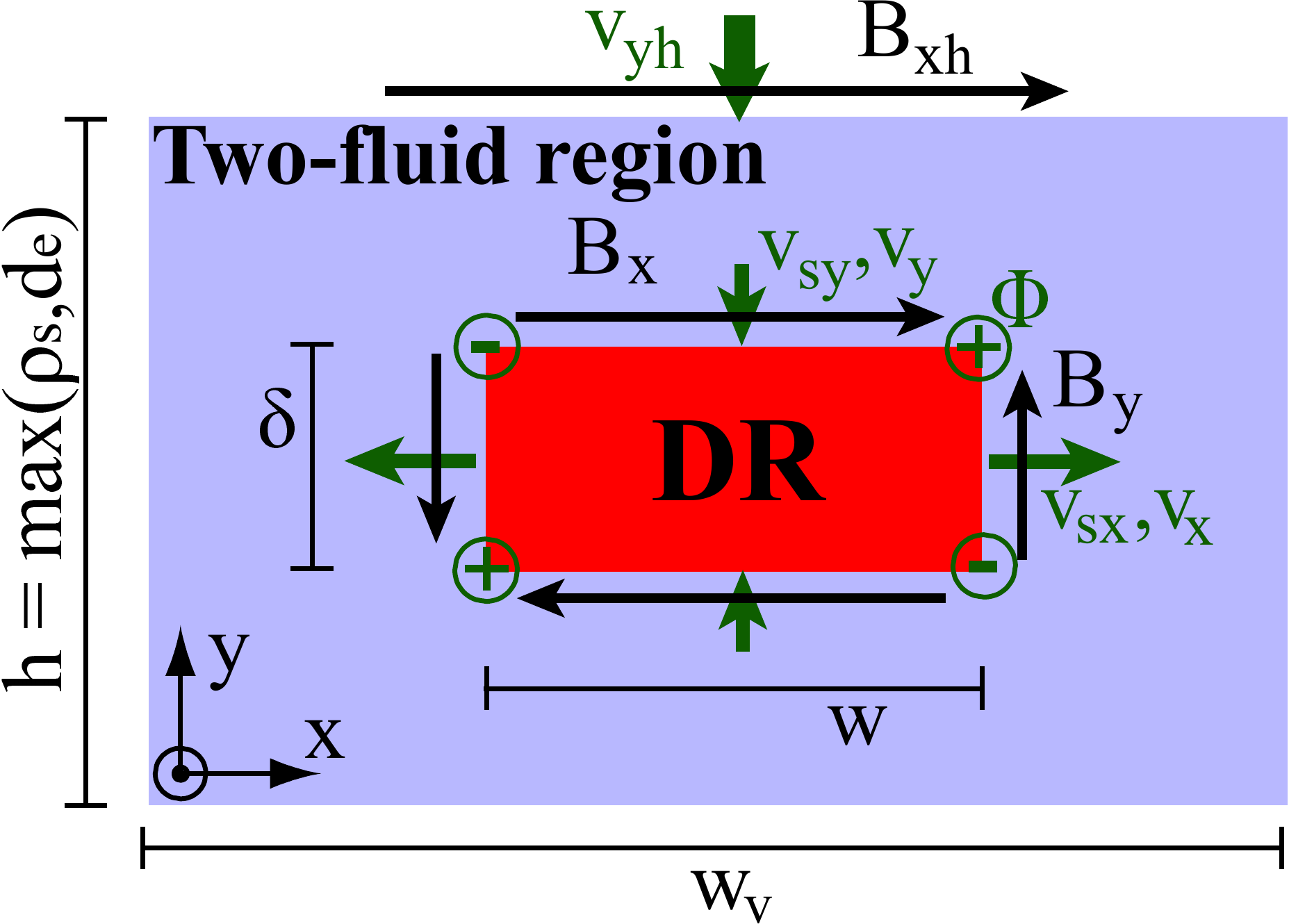}
\caption{\label{fig:dr}Dissipation region (DR) embedded in a larger region of thickness $h=\textrm{max}[\rho_s,d_e]$ at which two-fluid effects become important. The length $w$ is defined as the extent of the $v_{sx}$ outflow jets, and $w_v$ the extent of the $v_x$ outflow. Reprinted with permission from Stanier et al., Phys. Plasmas, 22, 010701, (2015). Copyright (2015) American Institute of Physics.}
\end{figure}

To construct the analytic model, Eqs.~(\ref{vorteqn}) and (\ref{vecBeqn}) are discretised at the DR as shown in Fig.~\ref{fig:dr}, using the technique of Refs.~\onlinecite{chacon07,chacon08,simakov10}. The discrete magnetic field components are defined on the edges of a rectangular DR of thickness $\delta$ and length $w$, so that $B_x = \boldsymbol{\hat{x}} \cdot \boldsymbol{B}(0,\delta/2)$, $B_y = \boldsymbol{\hat{y}} \cdot \boldsymbol{B}(w/2,0)$, and the discrete flow stream function is $\Phi = -\phi(w/2,\delta/2)$. Then, the inflow and outflow velocities are given by $V_x = 2\Phi/\delta$, $V_y = -2\Phi/w$. It is assumed that the DR is quasi-steady, so that derivatives with respect to time are small compared with other terms and can be neglected. This is normally true when the system reaches non-linear saturation, such as at the time of peak reconnection rate in the island-coalescence problem as shown in Sec.~\ref{sec:numresults} and Ref.~\onlinecite{simakov06}. Also, neglecting numerical factors of order unity gives three equations for the five unknowns $(\delta,w, B_x,B_y, \Phi)$ 
\begin{equation}\label{discvorteqn}{\frac{\Phi^2}{\delta w}}{\left(\frac{1}{\delta^2}{-}\frac{1}{w^2}\right)}{+}{\left(\frac{B_x}{w}{+}\frac{B_y}{\delta}\right)}{\left(\frac{B_y}{w}{-}\frac{B_x}{\delta}\right)}{=}{-\mu\Phi}{\Delta^2},\end{equation}
\begin{equation}\label{discbxeqn}{-\frac{\Phi}{\delta w} }{\left[B_x\left(1{+}\rho_s^2 \Delta\right){-}d_e^2 {\left(\frac{B_y}{\delta w}{-}\frac{B_x}{\delta^2}\right)}\right]}{=}{D}{\left(\frac{B_y}{\delta w}{-}\frac{B_x}{\delta^2}\right)},\end{equation}
\begin{equation}\label{discbyeqn}{\frac{\Phi}{\delta w}}{\left[B_y\left(1{+}\rho_s^2\Delta\right){-}d_e^2 {\left(\frac{B_x}{\delta w}{-}\frac{B_y}{w^2}\right)}\right]}{=}{D}{\left(\frac{B_x}{\delta w}{-}\frac{B_y}{w^2}\right)},\end{equation}
where $D \,{=}\, \eta \,{+}\, \eta_H\Delta$, and $\Delta \,{=}\, \delta^{-2} \,{+}\, w^{-2}$. This set of discrete equations combine the finite-$\rho_s$ terms in Ref.~\onlinecite{simakov10} and the finite-$d_e$ terms from the discrete equations in Ref.~\onlinecite{chacon08}. Eqs.~(\ref{discvorteqn})-(\ref{discbyeqn}) are invariant under plasma flow reversal about the DR $(B_x,B_y,\Phi,\delta) \leftrightarrow  (B_y,B_x,-\Phi,w)$. 

Defining $\xi = \delta/w$, $b=B_y/B_x$, $\hat{\rho}_s = \rho_s/\delta$ and $\hat{d}_e = d_e/\delta$, Eqs.~(\ref{discbxeqn},\ref{discbyeqn}) give
\begin{equation}\label{bxieqn}\frac{b}{\xi} = \frac{1 + \hat{\rho}_s^2\left(1+\xi^2\right) + 2\hat{d}_e^2}{1+\hat{\rho}_s^2\left(1+\xi^2\right)+2\hat{d}_e^2\xi^2},\end{equation}
which, after substitution back into Eq.~(\ref{discbyeqn}), gives an equation for the discrete stream function
\begin{equation}\label{phieqn}\Phi = \frac{D\left(\xi^{-1} - \xi\right)}{1 + \big(\hat{\rho}_s^2 + \hat{d}_e^2\big)\big(1+\xi^2\big)}.\end{equation}
Without loss of generality, we consider $\Phi > 0$ so that $\xi < 1$. Substituting Eqs.~(\ref{bxieqn}) and (\ref{phieqn}) into Eq.~(\ref{discvorteqn}) gives the master equation for $\xi$
\begin{eqnarray}\label{master1}\frac{\left[1 + \hat{\rho}_s^2\left(1 + \xi^2\right)\right]\left[1+\left(\hat{\rho}_s^2+\hat{d}_e^2\right)\left(1+\xi^2\right)\right]^3}{\left[1+\hat{\rho}_s^2\left(1+\xi^2\right)+2\hat{d}_e^2\xi^2\right]^2} = \nonumber\\ 
\frac{\left(1-\xi^2\right)^2}{S^2 \xi^4} + \frac{\left(1+\xi^2\right)^2 \left[1+\left(\hat{\rho}_s^2+\hat{d}_e^2\right)\left(1+\xi^2\right)\right]}{S S_\mu \xi^4},\end{eqnarray}
where $S=\sqrt{2}B_x w/D$ is the Lundquist number and $S_\mu = \sqrt{2} B_x w/\mu$. This equation can be solved numerically for $\xi$ (and therefore $\delta$) for given parameters $B_x$ and $w$. The reconnection rate ($E_z$) in steady-state ($\partial_t j = 0$) can then be calculated using the discrete expression
\begin{equation}\label{ezeqn1}E_z \approx D\left(\frac{B_x}{\delta} - \frac{B_y}{w}\right) \approx \frac{\sqrt{2} B_x^2 S^{-1}\left(\xi^{-1} - \xi\right)}{1 + 2\hat{d}_e^2\xi^2/\left[1+\hat{\rho}_s^2\left(1+\xi^2\right)\right]}.\end{equation}
Here we solve for $\xi$ and $E_z$ analytically, noting that large $E_z$ preferentially occur for $\xi \ll 1$. This motivates the approximation $1 + \xi^2 \approx 1 - \xi^2 \approx 1$, which considerably simplifies the above expressions. We also assume that $2\hat{d}_e^2\xi^2  = 2d_e^2/w^2 \ll 1$. The validity of these two assumptions was checked for the numerical simulations presented in this paper. Finally, in this paper, we concentrate on viscous DRs  ($\eta \,{=}\, 0$; $\eta_H,\mu \,{\neq}\, 0$), as resistivity alone can not prevent the DR collapsing to zero thickness once $\delta < h$, where  $h=\textrm{max}[\rho_s,d_e]$.~\cite{zocco09,simakov10} The resulting equation for $\delta (B_x,w)$ is given by
\begin{equation}\label{master2}{\frac{\delta^4}{w^8}}{\frac{\left(\delta^2\,{+}\,\rho_s^2\,{+}\,d_e^2\right)^3}{\delta^2\,{+}\,\rho_s^2}}\,{=}\,
\frac{1}{S_H^2}{\left[1\,{+}\,\frac{\mu (\delta^2\,{+}\,\rho_s^2\,{+}\,d_e^2)}{\eta_H}\right]},\end{equation}
where $S_H\,{\equiv}\,\sqrt{2} B_x w^3/\eta_H$ is the hyper-resistive Lundquist number, and the quasi-steady ($\partial_t j = 0$) reconnection rate is
\begin{equation}\label{rateeqn}E_z \approx D\left(\frac{B_x}{\delta} - \frac{B_y}{w}\right)\approx \eta_H B_x/\delta^3.\end{equation} 

\subsection{Single fluid scalings}
We first note that the usual resistive single-fluid scalings,~\cite{sweet58,parker57} including the correction for finite ion-viscosity,~\cite{park84} can be straightforwardly obtained from Eqs.~(\ref{master1},~\ref{ezeqn1}) as was shown in Ref.~\onlinecite{simakov10}. Concentrating here on hyper-resistive layers with Eqs.~(\ref{master2},~\ref{rateeqn}), and taking the single-fluid limit $(\delta^2 \gg d_e^2,\rho_s^2)$ with zero ion-viscosity ($\mu=0$), gives
\begin{equation}\label{slowscalings}\delta = \delta_H \equiv w\,S_H^{-1/4}, \quad E_z = E_{zH} \equiv \sqrt{2} B_x^2 S_H^{-1/4},\end{equation}
which are the expected single-fluid hyper-resistive scalings for the DR thickness and reconnection rate (see e.g. Ref.~\onlinecite{huang13}). Here, as $\rho_s = 0$, there is no separation between in-plane ion and electron flows, $\boldsymbol{v} = \boldsymbol{v}_s$, so that the DR length $w$ is equal to the extent of the single-fluid outflow jets, $w_v$, see Fig.~\ref{fig:dr}. These scalings can also be generalised for finite ion-viscosity.~\cite{stanier15}

\subsection{\label{sec:nofdwscalings}Scalings for case without FDWs}
In what follows, we assume for simplicity that the ion-viscosity is set to a fixed value of $\mu/\eta_H = 1/(\rho_s^2 + d_e^2)$. However, the following two-fluid scalings can be straightforwardly generalised to relax this assumption.

We consider first the two-fluid case without FDWs, with $d_e^2 \gg \rho_s^2, \delta^2$. In the limit of small-$\rho_s$, such that $d_e^2 \gg \delta^2 \gg \rho_s^2$, Eqs.~(\ref{master2},~\ref{rateeqn}) give the DR thickness and reconnection rate as
\begin{equation}\label{inertialdelta}\delta = \frac{\eta_Hw}{B_x d_e^3},\quad E_z = \frac{B_x^4 d_e^9}{\eta_H^2 w^3}.\end{equation}
The reconnection rate appears to be ``super-fast", that is $E_z \,{\propto}\, \eta_H^{\alpha}$ with $\alpha \,{<}\, 0$. However, the free parameters $B_x$ and $w$ have not yet been specified, and they do scale with $\eta_H$ as will be described.

The expression for the DR thickness in Eq.~(\ref{inertialdelta}) is similar to that for a low-$\beta$ electron-positron plasma, $\delta {=} \mu w/(B_x d_e)$, in Ref.\onlinecite{chacon08}. Indeed, for the electron-positron mass ratio, ${m_i {=} m_e}$, such that $\eta_H {=} d_i^2 \mu_e {=} d_e^2 \mu$, we obtain the electron-positron result. This is because the two-field Eqs.~(\ref{vorteqn},\ref{vecBeqn}) reduce, in the limit $m_i{=}m_e$ and $\rho_s{=}0$, to the low-$\beta$ electron-positron plasma equations studied in Ref.~\onlinecite{chacon08}, apart from numerical factors of order unity. 

As discussed in Ref.~\onlinecite{chacon08}, the electron viscous (hyper-resistive) DR supports the reconnection electric field at the X-point in steady-state, but the bulk of the current is supported at $d_e$-scale such that the magnetic field at the DR edge scales like $B_x \approx \delta B_{xd}/d_e$, where $B_{xd}\, {=}\, \boldsymbol{\hat{x}}\,{\cdot}\, \boldsymbol{B}(0,d_e/2)$ is shown in Fig.~\ref{fig:dr}. Finally, as $\rho_s$ is small in this limit, there is little separation of ion and electron in-plane flows, and so $w\approx w_v$. This gives the following scalings for the DR thickness and reconnection rate
\begin{equation}\label{descalings}\delta = \frac{\delta_d^2}{d_e}, \quad E_z = B_{xd}^2 \frac{d_e}{w_v},\end{equation}
where $\delta_d = w_v (\eta_H/B_{xd} w_v^3)^{1/4}$ is equal to the single-fluid thickness in Eq.~(\ref{slowscalings}), evaluated with $B_{xd}$ and $w_v$. $E_z$ has no explicit dependence on hyper-resistivity in Eq.~(\ref{descalings}), but it is still a function of the parameters $B_{xd}$ and $w_v$ which may have some $\eta_H$ dependence. It is shown in Section~\ref{sec:numresults} that this rate is indeed dissipation independent as these parameters do not depend on $\eta_H$, in agreement with the electron-positron result.~\cite{chacon08} 

In the second limit $d_e^2 \gg \rho_s^2 \gg \delta^2$, Eqs.~(\ref{master2},~\ref{rateeqn}) give the DR thickness and reconnection rate as
\begin{equation}\label{descalingsrhos} \delta =  \left(\frac{w \eta_H \rho_s}{B_x d_e^3}\right)^{1/2},\quad E_z = \left(\frac{B_x^5 d_e^9}{\eta_H \rho_s^3 w^3}\right)^{1/2}.\end{equation}
The rate again would be super-fast if the free-parameters $B_x, w \propto \eta_H^0$. However, the rate can not become singular as $\eta_H \rightarrow 0$,  so we expect it to remain dissipation-independent when finite-$\rho_s$ effects become important, provided that the DR has the capacity to self-adjust $B_x$ and $w$ appropriately. In this limit, the current density is still supported at $d_e$-scale, so $B_x$ scales as before. Comparing Eqs.~(\ref{descalingsrhos},~\ref{descalings}), the DR length $w$ must adjust as $w = (\delta/\rho_s) w_v$ to keep the same rate. In Section~\ref{sec:numresults}, we verify that the DR does have the capacity to self-adjust in this manner, even in the absence of fast-dispersive waves, to give the DR thickness and dissipation-independent rate of Eq.~(\ref{descalings}) that is constant across both limits.

\subsection{\label{sec:fdwscalings}Scalings for case with FDWs}

Next we consider the two-fluid case with FDWs in two limits; $\rho_s^2 \gg d_e^2 \gg \delta^2$, and $\rho_s^2 \gg \delta^2 \gg d_e^2$. In both limits, the DR thickness and reconnection rate are found from Eqs.~(\ref{master2},~\ref{rateeqn}) as
\begin{equation}\label{initrhosscalings}\delta = \left(\frac{\eta_H w}{B_x \rho_s^2}\right)^{1/2},\quad E_z = \left(\frac{B_x^5 \rho_s^6}{\eta_H w^3}\right)^{1/2},\end{equation}
where again $E_z$ appears proportional to a negative power of $\eta_H$. Previous studies~\cite{kleva95,rogers01,bhattacharjee05,schmidt09,simakov10} have shown that reconnection is indeed dissipation independent in this regime, and so we expect that the DR is able to self-adjust to give a fast rate, as in the $d_e > \rho_s$ case. In the first limit $\delta < d_e$ so that $B_x \propto \delta B_{xd}/d_e$ as before, and a rate that is explicitly independent of $\eta_H$ requires that $w \propto \delta$. We show from numerical simulations, in Sec.~\ref{sec:numresults}, that the fixed-aspect ratio scaling $\delta/w \propto \rho_s/w_v$, from the $d_e^2 \gg \rho_s^2 \gg \delta^2$ limit, holds whenever $\delta < \rho_s$, regardless of the value of $d_e$.

The second limit has $\delta > d_e$, and so a scaling for $B_x$ is not known \textit{a priori} from physical arguments. However, with the $w = (\delta/\rho_s) w_v$ scaling for $\delta < \rho_s$, a dissipation independent rate in this limit also requires that $B_x \propto \delta$. From simulations in Sec.~\ref{sec:numresults} we find for both limits that, to a good approximation, $B_x \approx (\delta/\rho_s) B_{xs}$ with $B_{xs} = \boldsymbol{\hat{x}}\cdot \boldsymbol{B}(0,\rho_s/2)$. With this, the DR thickness and rate in both limits is given by
\begin{equation}\label{rhosscalings}\delta =  \frac{\delta_s^2}{\rho_s}, \quad E_z = B_{xs}^2 \frac{\rho_s}{w_v},\end{equation}
where $\delta_s = w_v (\eta_H/B_{xs} w_v^3)^{1/4}$. Again, this rate has no explicit dependence on $\eta_H$, and we verify in Sec.~\ref{sec:numresults} that the two-fluid region parameters $B_{xs}$ and $w_v$ have no $\eta_H$ dependence.

The DR thickness and reconnection rate in cases with and without FDWs can be expressed simply as
\begin{equation}\label{bothscalings}\delta =  \frac{\delta_h^2}{h}, \quad E_z = v_{yh}B_{xh} = B_{xh}^2 \frac{h}{w_v},\end{equation}
where $v_{yh}= B_{xh} h/w_v$ is the inflow velocity into the larger two-fluid region, see Fig.~\ref{fig:dr}, constrained by flow continuity with the outflow equal to the upstream Alfv\'en speed ($B_{xh}$ in these normalised units). This form of the reconnection rate can be arrived at from simple steady-state arguments, considering only the outer two-fluid region. But, by starting from a discrete model of the DR, we have illustrated how $\delta$, $B_x$ and $w$ must scale in each case to give this result. 

The form of the rate in Eq.~(\ref{bothscalings}) also holds for the single-fluid limit, when $h=\delta_H$ and $w=w_v$, simply because the reconnection rate must be the same as the upstream rate of inflowing flux in steady-state ($\partial_y E_z = - \partial_t B_x = 0$). However, the mechanism whereby this result is obtained differs between single and two-fluid regimes. In the single-fluid regime, the reconnection rate is limited by the DR, so that the upstream rate of inflowing flux adjusts and slows down to the DR rate in steady-state. This is in contrast to the two-fluid regime, where the rate is instead bounded by rate of flux inflow to the outer two-fluid region, and the DR self-adjusts to match this upstream rate. The crucial difference gives a rate explicitly dependent on $\eta_H$ in the single-fluid case, and a rate independent of DR physics in the two-fluid case.

\section{\label{sec:method}Numerical Method}

We solve implicitly the vorticity and magnetic-flux formulation of Eqs.~(\ref{vorteqn},~\ref{vecBeqn}) with a Jacobian-Free Newton Krylov (JFNK) based algorithm.~\cite{chacon02,chacon03,chacon14} To accelerate convergence, a physics-based preconditioning strategy was developed, where the hyperbolic couplings present due to the fast-dispersive wave terms in the preconditioner are parabolised before they are inverted using multigrid methods.~\cite{chacon03,chacon14} This preconditioning has no effect on the solution, but reduces the numerical stiffness when taking larger time steps than the explicit step associated with the fast-dispersive waves. The result is a large effective speed-up over explicit schemes.

The spatial operators in Eqs.~(\ref{vorteqn},~\ref{vecBeqn}) are discretised using second-order centred finite differences, apart from the advection terms, which are discretised using quadratic upstream interpolation (QUICK). For timestepping, we use either second-order Crank-Nicolson or Backward-Differentiation formula (BDF-2). We use grid-packing focused around the X-point to adequately resolve the DR, which is often sub-$d_e$ scale, in all simulations. With strict control of numerical dissipation in the scheme used, and no additional high-order dissipation than is stated explicitly in Eqs.~(\ref{vorteqn},~\ref{vecBeqn}), we are able to accurately measure how DR parameters such as the thickness, length and upstream magnetic field depend on the magnitude of the physical dissipation. Both grid and timestep convergence studies were performed for a selection of simulations to verify the accuracy of the results presented. Finally, we note that the solver successfully reproduces the linear tearing eigenmode structure and growth-rate~\cite{chacon14} in the two-fluid (large $\rho_s$ regime), as calculated by a separate linear stability code.

The simulations presented in this paper use one of two typical reconnection problem set-ups; the island coalescence problem, and the single Harris-sheet. The island coalescence simulations have initial flux-function
\begin{equation}\label{initialcondic}\psi_{ic} = \lambda \ln{\left[\cosh{(x/\lambda)} + \epsilon \cos{(y/\lambda)}\right]},\end{equation}
with $\lambda = 1/(2 \pi)$, $\epsilon = 0.2$, and with a sinusoidal perturbation of $0.1\%$. This problem is solved in a quarter box $(x,y) \in [0,1]\times [0,1]$ with symmetry (or anti-symmetry, depending on the variable) boundaries at $x=0$, $y=0$ and $y=1$, and a perfectly conducting boundary at $x=1$.

The Harris-sheet runs use the flux-function 
\begin{equation}\label{initialcondha}\psi_{Ha} = -\lambda \ln{\left[\cosh{(y/\lambda)}\right]}.\end{equation}
They are run in a half-box $(x,y)\in [-L_x,L_x]\times[0,0.5]$, with equilibrium current sheet thickness $\lambda = 1/(8\pi)$, and initiated with a sinusoidal perturbation with magnitude equal to $3\%$ of the upstream boundary magnetic field strength. The half-box domain ensures that any secondary islands that form at $x=0$ are not trapped indefinitely.

\section{\label{sec:numresults}Dissipation independence}
\subsection{Reconnection rates}

\begin{figure}
\includegraphics[scale=0.2]{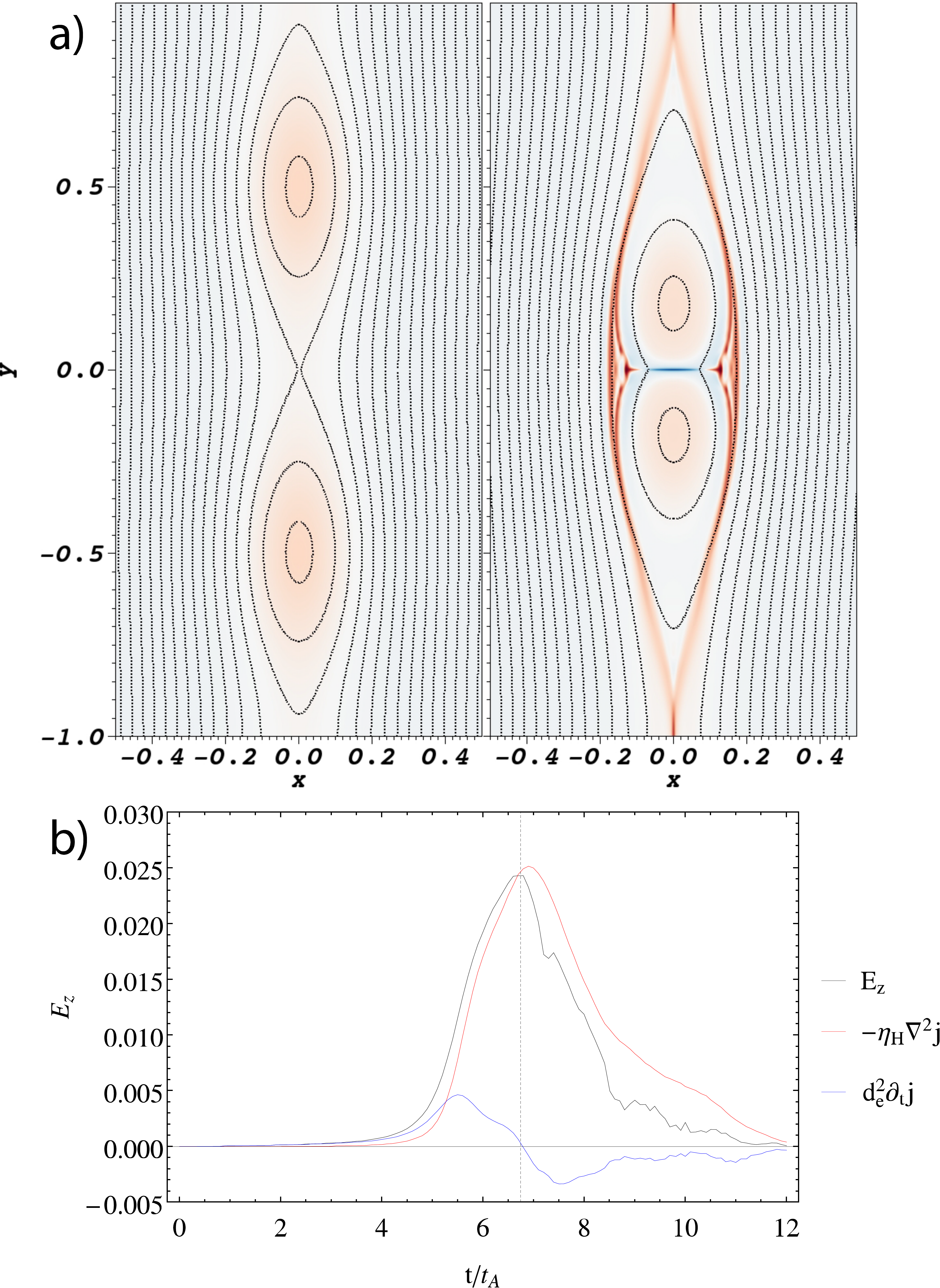}
\caption{\label{fig:ic-setup} a) Current density (colour scale) and magnetic flux contours (black dotted) for an island coalescence simulation with $d_e = 0.01$, $\rho_s = 0.002$, $\eta_H=10^{-9}$, $\mu=10^{-5}$ at $t=0$ (left) and at peak rate ($t=6.8$, right). The quarter-domain simulation has been reflected in $x=0$ and $y=0$, and then cropped at $|x|=0.5$. b) Reconnection rate ($E_z$, black) against time for this simulation, with inertial (blue) and hyper-resistive (red) contributions. The vertical dashed line indicates the time of peak rate.}
\end{figure}

Figure~\ref{fig:ic-setup}a shows the flux and current profiles in the initial conditions ($t=0$, left), and at the time of peak reconnection rate ($t=6.8$, right) for an island coalescence run with $d_e = 0.01$, $\rho_s = 0.002$ (without FDWs), $\eta_H=10^{-9}$ and $\mu = 10^{-5}$. At the time of peak rate, the X-point has collapsed and a current-sheet (blue) has formed between the two islands. The reconnection rate is plotted against time in Fig.~\ref{fig:ic-setup}b, where it has been normalised by the Alfv\'enic rate at the upstream boundary $(x=1)$. The initially ideal island coalescence instability quickly reaches non-linear saturation when the rate peaks, at which time-dependent terms such as the electron inertial contribution to the X-point electric field (blue) are typically small. For this reason, we use island coalescence simulations to benchmark the quasi-steady discrete model of the DR presented in Section~\ref{sec:discretemodel}.

\begin{figure}
\includegraphics[scale=0.65]{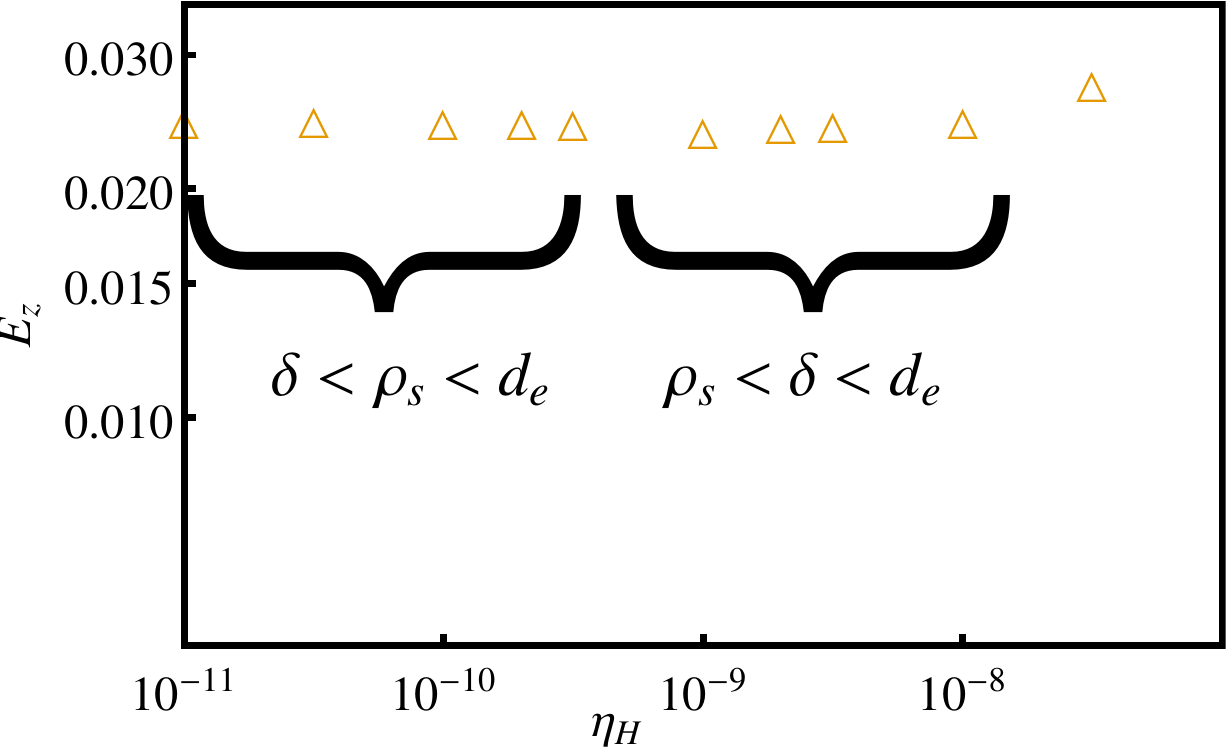}
\caption{\label{fig:rate-nofdw} Peak reconnection rate against hyper-resistivity from island coalescence simulations with $d_e = 0.01$, $\rho_s = 0.002$ and $\mu/\eta_H = 10^4$. The relative size of the DR thickness $\delta$ with respect to $d_e$ and $\rho_s$ at the time of peak rate is indicated.}
\end{figure}

Figure~\ref{fig:rate-nofdw} shows the peak rate from a number of simulations that are identical to that shown in Fig.~\ref{fig:ic-setup}, but with different hyper-resistivity and ion-viscosity. For this scaling study, we keep the ratio of ion-viscosity to hyper-resistivity fixed as $\mu/\eta_H =1/(\rho_s^2 + d_e^2) \approx 10^4$ to compare with the scalings presented in Sec.~\ref{sec:discretemodel}. As hyper-resistivity is reduced between simulations, the DR thickness $\delta$, measured at the peak rate, decreases so that it falls below $d_e$-scale for $\eta_H \leq 10^{-8}$ and also falls below $\rho_s$ for $\eta_H \leq 10^{-9.5}$. It is clear for this case without FDWs that the reconnection rate is independent of the dissipation, $\eta_H$, provided that $\delta < d_e$. In addition, there is no change when $\delta$ falls below the smaller of the two-fluid scales, $\rho_s$, as anticipated in Sec.~\ref{sec:nofdwscalings}.

\begin{figure}
\includegraphics[scale=0.65]{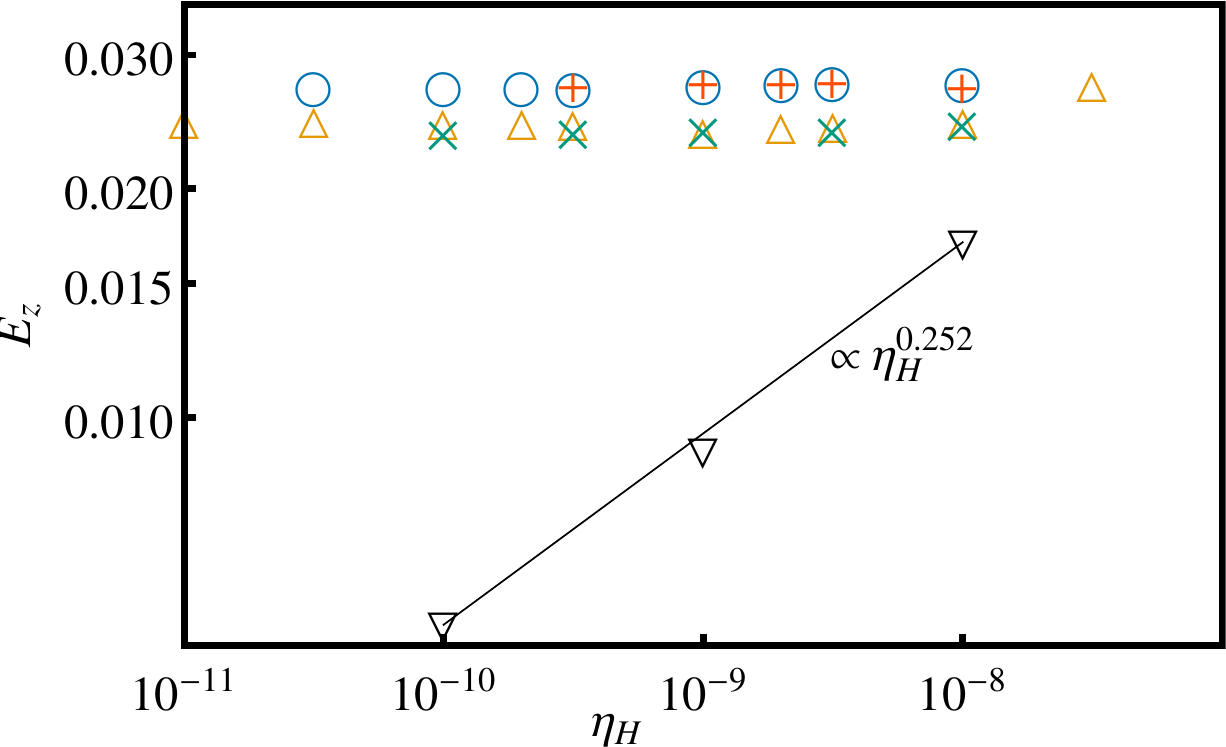}
\caption{\label{fig:ic-rates} Peak rates from island coalescence runs with $d_e = 5\rho_s = 0.01$ (orange $\triangle$), $d_e = 0.01, \,\rho_s = 0$ (green $\times$), $\rho_s = 5d_e = 0.01$ (blue $\bigcirc$), $\rho_s = 0.01,\,d_e = 0$ (red $+$), and $\rho_s=d_e=0$ (black $\bigtriangledown$). Reprinted with permission from Stanier et al., Phys. Plasmas, 22, 010701, (2015). Copyright (2015) American Institute of Physics. }
\end{figure}

Figure~\ref{fig:ic-rates} also shows the peak reconnection rates against hyper-resistivity from a number of island coalescence simulations. Shown are two-fluid cases without FDWs (orange $\triangle$ for $d_e = 0.01$ and $\rho_s = 0.002$, which are the same as in Fig.~\ref{fig:rate-nofdw}, and green $\times$ for $d_e = 0.01$ and $\rho_s = 0$), with FDWs (blue $\bigcirc$ for $\rho_s = 0.01$ and $d_e = 0.002$, and red $+$ for $\rho_s = 0.01$ and $d_e = 0$), and single-fluid runs (black $\bigtriangledown$, where $\rho_s = d_e = 0$). The single-fluid rate depends on dissipation as $E_z \propto \eta_H^{0.25}$, which is given in Eq.~(\ref{slowscalings}) with $B_x, w \propto \eta_H^0$ in the single-fluid regime. However, all two-fluid runs with and without FDWs are independent of $\eta_H$. Also, there is no appreciable change in the rate between $d_e = 0.01$ simulations (orange $\triangle$ and green $\times$) and $\rho_s = 0.01$ simulations (blue $\bigcirc$ and red $+$), and the rate is also unchanged when the smaller of the two-fluid scales is set to zero (green $\times$ and red $+$). This is in agreement with the scaling of Eq.~(\ref{bothscalings}), where only the larger of the two-fluid scales $h=\textrm{max}[\rho_s,d_e]$ is explicit in the reconnection rate.  

\begin{figure}
\includegraphics[scale=0.23]{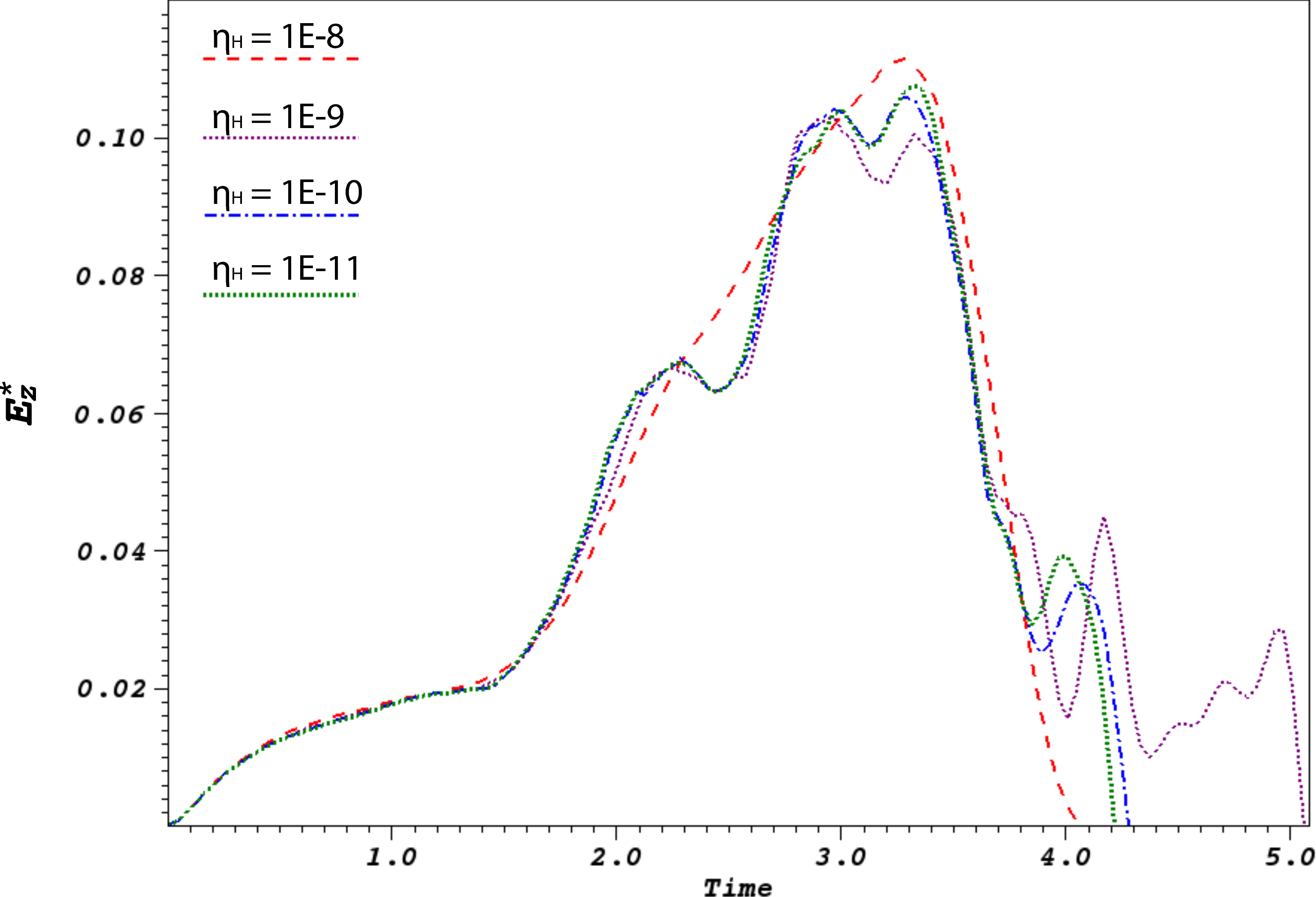}
\caption{\label{fig:harris-dis}Normalised reconnection rate $E_z^*$ vs. time for Harris-sheet simulations with $d_e = 0.01$, $\rho_s = 0.002$ for different hyper-resistivity. }
\end{figure}

All simulations discussed so far have used the island coalescence problem set-up. Fig.~\ref{fig:harris-dis} shows the reconnection rate against time for Harris sheet simulations with $d_e = 0.01$, $\rho_s = 0.002$ for different $\eta_H$. Here, the rate is measured as
\begin{equation}\label{harrisrate}E_z^* = \frac{1}{(\boldsymbol{\hat{x}}\cdot\boldsymbol{v}_A)(\boldsymbol{\hat{x}}\cdot\boldsymbol{B})} \frac{\partial \psi_r}{\partial t},\end{equation}
where $\psi_r\,{=}\,\textrm{max}(\psi)\,{-}\,\textrm{min}(\psi)$ is the flux difference between the X and O-points, and $(\boldsymbol{\hat{x}}\cdot\boldsymbol{v}_A)(\boldsymbol{\hat{x}}\cdot\boldsymbol{B})$ is the Alfv\'enic rate at $4d_e$ upstream of the main X-point. In the series of simulations shown in Fig.~\ref{fig:harris-dis}, there is no secondary island formation. The reconnection rate is again independent of $\eta_H$, suggesting that dissipation independence is a general property of reconnection in the two-fluid regime without FDWs, rather than being specific to the island coalescence problem set-up.

\subsection{Discrete model scalings}

The scalings for $B_x$ and $w$ given in Sec~\ref{sec:discretemodel} are now verified using the previously described island coalescence simulations. For each simulation, we measure $(\delta, w, B_x, w_v,B_{xh})$ at peak reconnection rate. The thickness, $\delta$, is measured in the same way as in Refs.~\onlinecite{chacon08} and~\onlinecite{stanier15}, which is appropriate for viscous DRs. The lengths $w$ and $w_v$, shown in Fig.~\ref{fig:dr}, are defined as the distance between the maxima of $|\boldsymbol{\hat{x}}\cdot \boldsymbol{v}_s|_{y=0}$ and $|\boldsymbol{\hat{x}}\cdot \boldsymbol{v}|_{y=0}$, respectively. The magnetic field strengths $B_x$ and $B_{xh}$ are evaluated at $(x,y) = (0,\delta/2)$ and $(x,y) = (0,h/2)$ respectively, as shown in Fig.~\ref{fig:dr}. 

\begin{figure}
\includegraphics[width=0.4\textwidth]{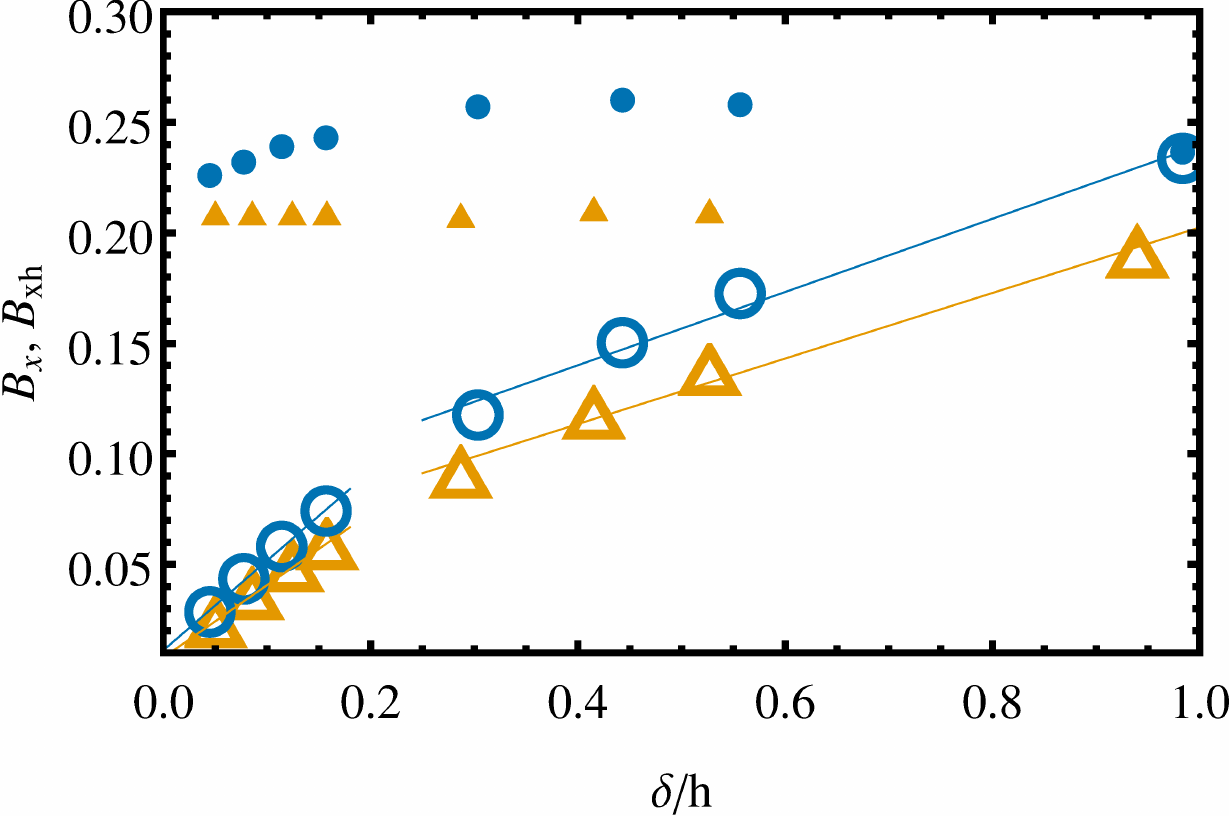}
\caption{\label{fig:scalings-bx}Scaling of $B_x$ (hollow markers with linear fits) and $B_{xh}$ (small filled markers) against $\delta/h$, where $h=\textrm{max}[\rho_s,d_e]$, for $d_e = 5\rho_s = 0.01$ (orange $\triangle$) and $\rho_s = 5d_e = 0.01$ (blue $\bigcirc$). Reprinted with permission from Stanier et al., Phys. Plasmas, 22, 010701, (2015). Copyright (2015) American Institute of Physics.}
\end{figure}

In the discrete model scalings of Sec.~\ref{sec:discretemodel}, the scaling $B_x \propto (\delta/d_e) B_{xd}$ was used for $\delta < d_e$, as the current is supported at $d_e$-scale and magnetic field can not pile-up on scales below this. Also, when $\rho_s^2 \gg \delta^2 \gg d_e^2$, it was required that $B_x \propto \delta$ to give a dissipation-independent rate. Fig.~\ref{fig:scalings-bx} shows how $B_x$, and the magnetic field at the edge of the larger two-fluid region $B_{xh}$ (see Fig.~\ref{fig:dr}) scale with $\delta/h$, where $h=\textrm{max}[\rho_s,d_e]$. The $B_x$ parameter decreases with $\delta/h$, compared with $B_{xh}$ that is approximately constant over the full range of $\delta/h$. Furthermore, the trends of $B_x$ and $B_{xh}$ and the absolute values are similar in cases with and without FDWs. Here, the DR parameter $B_x$ has been fit with two straight lines for both cases, with and without FDWs, as there is a change in slope by $\approx 2$ when $\delta$ falls below the smaller of the two-fluid scales. This change in slope is not captured by the discrete model described in Sec.~\ref{sec:discretemodel}, as numerical factors of order unity are not retained in the discretisation of the two-field equations. 

\begin{figure}
\includegraphics[width=0.4\textwidth]{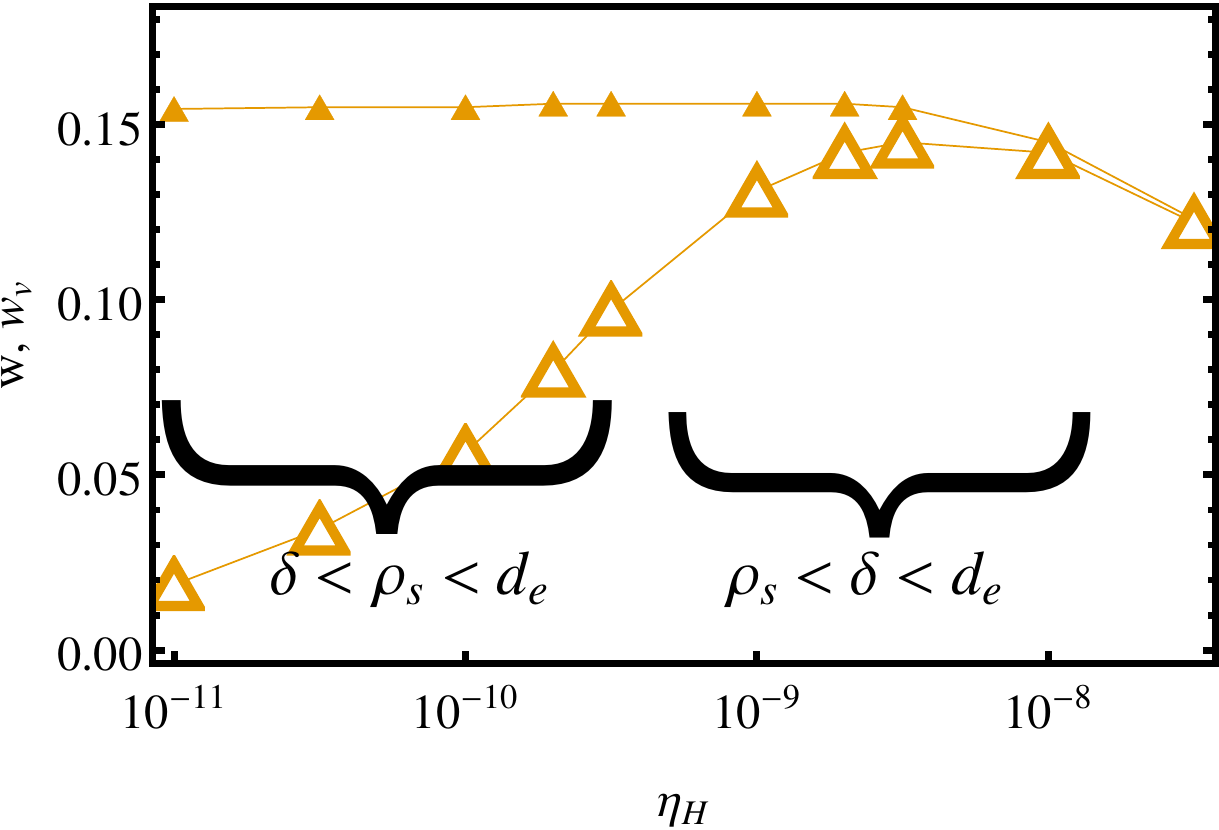}
\caption{\label{fig:nofdw-w}Scaling of $w$ (larger hollow marker) and $w_v$ (smaller filled marker) against hyper-resistivity for runs with $d_e = 5\rho_s = 0.01$ (orange $\triangle$). The relative size of the DR thickness $\delta$ with respect to $d_e$ and $\rho_s$ at the time of peak rate is indicated.}
\end{figure}

In the limit $d_e^2 \gg \rho_s^2 \gg \delta^2$ of the case without FDWs, it was argued that the DR length $w$ must scale with constant aspect-ratio as $w \propto (\delta/\rho_s) w_v$ when $\delta < \rho_s$, to give a constant and dissipation independent rate across the limits $d_e^2 \gg \delta^2 \gg \rho_s^2$ and $d_e^2 \gg \rho_s^2 \gg \delta^2$. Fig.~\ref{fig:nofdw-w} shows $w$ and $w_v$ at the time of peak reconnection rate plotted against $\eta_H$ for the island coalescence simulations with $d_e = 0.01$ and $\rho_s=0.002$ (the case without FDWs that is shown in Fig.~\ref{fig:rate-nofdw}). For $\rho_s < \delta < d_e$ the DR length $w$ is approximately $w \approx w_v$, as might be expected from the definition of the flux-carrying velocity, $\boldsymbol{v}_s$. However, for $\delta < \rho_s < d_e$, where $\rho_s$ effects become important, $w$ begins to shrink with respect to $w_v$. Across both limits $w_v$ is independent of $\eta_H$, as is required for a dissipation independent reconnection rate in Eq.~(\ref{descalings}). 

\begin{figure}
\includegraphics[width=0.4\textwidth]{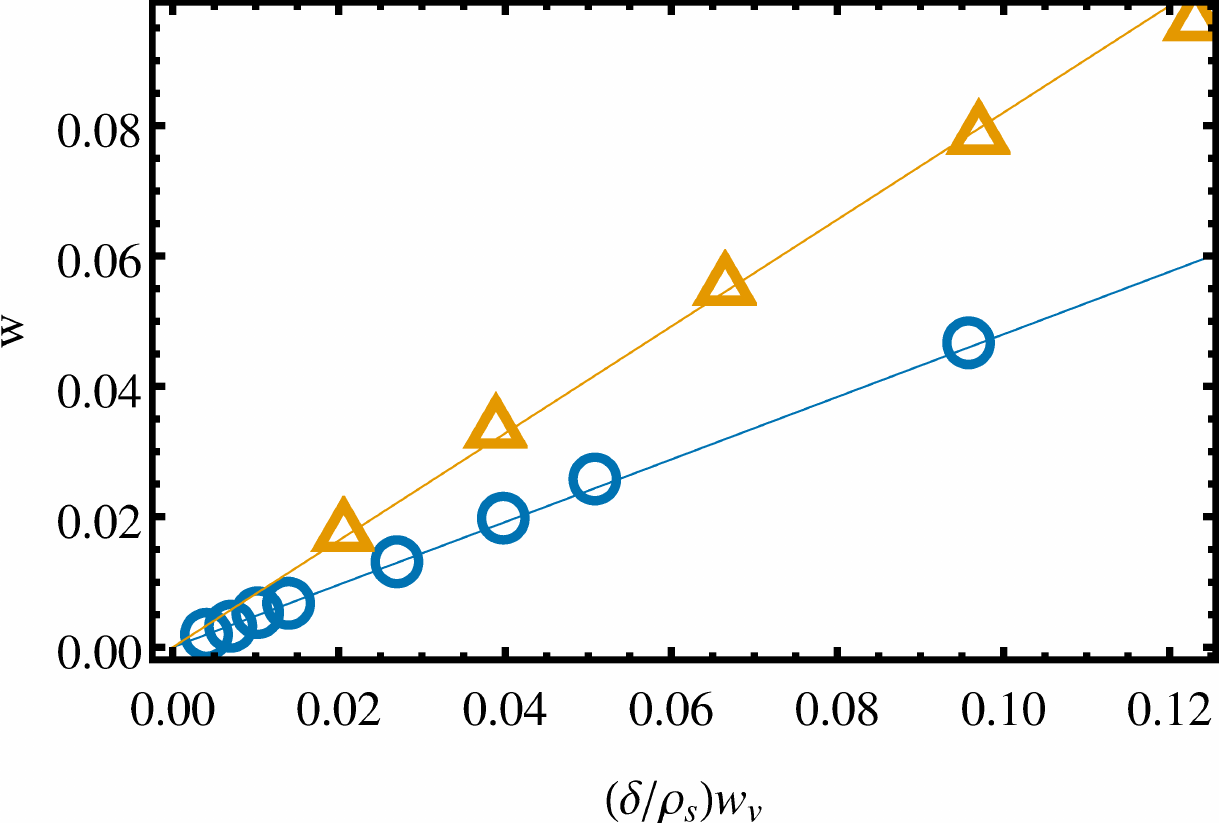}
\caption{\label{fig:wscalings}Scaling of $w$ against $(\delta/\rho_s) w_v$ from $d_e = 5\rho_s = 0.01$ runs with $\delta < \rho_s$ (orange $\triangle$), and from all runs with $\rho_s = 5d_e = 0.01$ (blue $\bigcirc$).}
\end{figure}

Figure~\ref{fig:wscalings} shows the values of $w$ from the simulations with $\delta < \rho_s < d_e$ in Fig.~\ref{fig:nofdw-w} plotted against $(\delta/\rho_s) w_v$. There is a clear linear scaling of $w$ with $\delta$ (quadratic in $\eta_H$ for Fig.~\ref{fig:nofdw-w}), which verifies that the DR self-adjusts in the manner required to maintain a constant and dissipation independent rate between the limits $d_e^2 \gg \delta^2 \gg \rho_s^2$ and $d_e^2 \gg \rho_s^2 \gg \delta^2$ (see Sec.~\ref{sec:nofdwscalings}). Also shown in Fig.~\ref{fig:wscalings} is the constant aspect-ratio DR scaling of $w \propto (\delta/\rho_s) w_v$ from the case with FDWs ($\rho_s = 0.01$, $d_e = 0.002$) that was used in  Sec.~\ref{sec:fdwscalings}. Here, there is a small change $(\approx \sqrt{2})$ in the slope of the linear fit between the $d_e^2 \gg \rho_s^2$ (no FDWs) and $\rho_s^2 \gg d_e^2$ (FDWs) cases that is not captured in our simple analytic model, as we lose numerical factors of order unity in the discretisation process.

\begin{figure}
\includegraphics[width=0.4\textwidth]{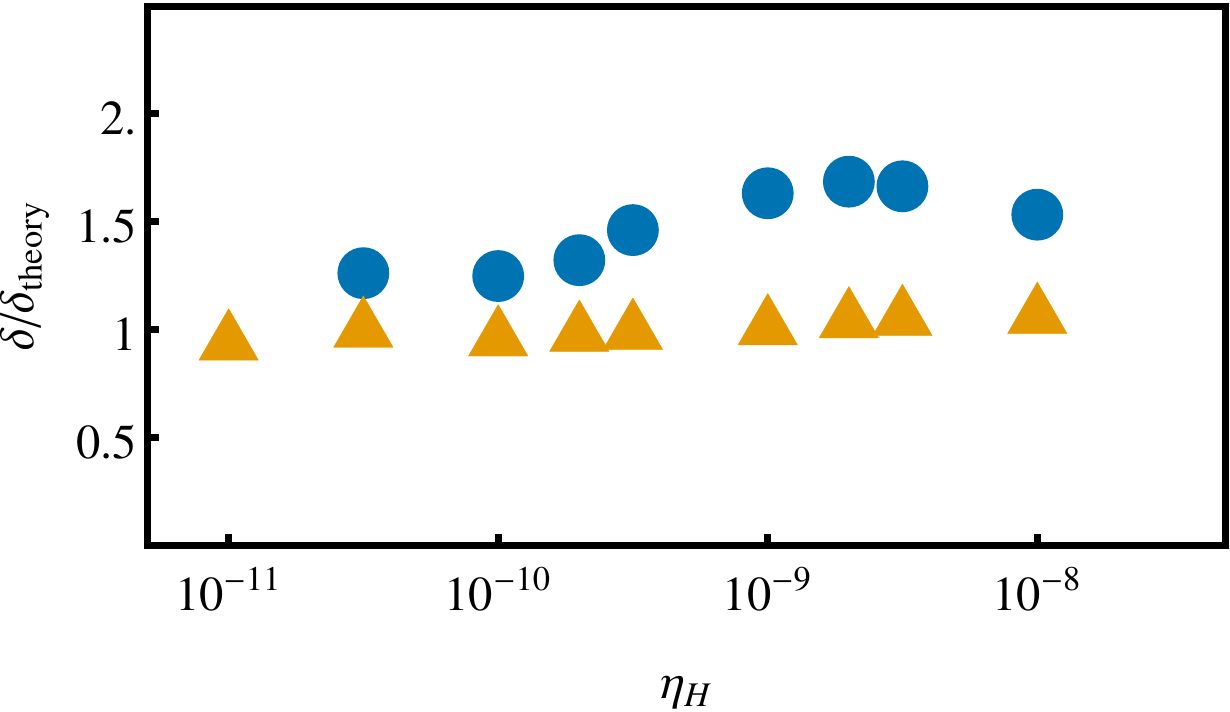}
\caption{\label{fig:deltavsth}Ratio of measured DR thickness $\delta$ to that from discrete model in Eq.~(\ref{bothscalings}) for island coalescence simulations with $d_e = 5\rho_s =0.01$ (orange $\triangle$) and $\rho_s = 5d_e = 0.01$ (blue $\bigcirc$). Reprinted with permission from Stanier et al., Phys. Plasmas, 22, 010701, (2015). Copyright (2015) American Institute of Physics.}
\end{figure}

Finally, Fig.~\ref{fig:deltavsth} shows a comparison of the measured DR thickness $\delta$ with the theoretical scaling of Eq.~(\ref{bothscalings}) for the cases with and without FDWs. The theoretical and measured values agree within a factor of $\approx 1.5$ over almost four orders-of-magnitude in $\eta_H$, or almost two orders-of-magnitude in $\delta$ (as $\delta \propto \eta_H^{1/2}$).

As was discussed in Sec.~\ref{sec:nofdwscalings}, the self-adjustment of the DR appears necessary to prevent unphysical ``super-fast" rates. Here, we have verified that the DR self-adjusts in the manner required to match the rate of flux inflow to the larger two-fluid region, which is an upper bound on the possible rate, rather than adjusting to give a slow (DR limited) rate. 

\subsection{\label{sec:comparekinetic}Comparison with kinetics}

\begin{figure}
\includegraphics[width=0.5\textwidth]{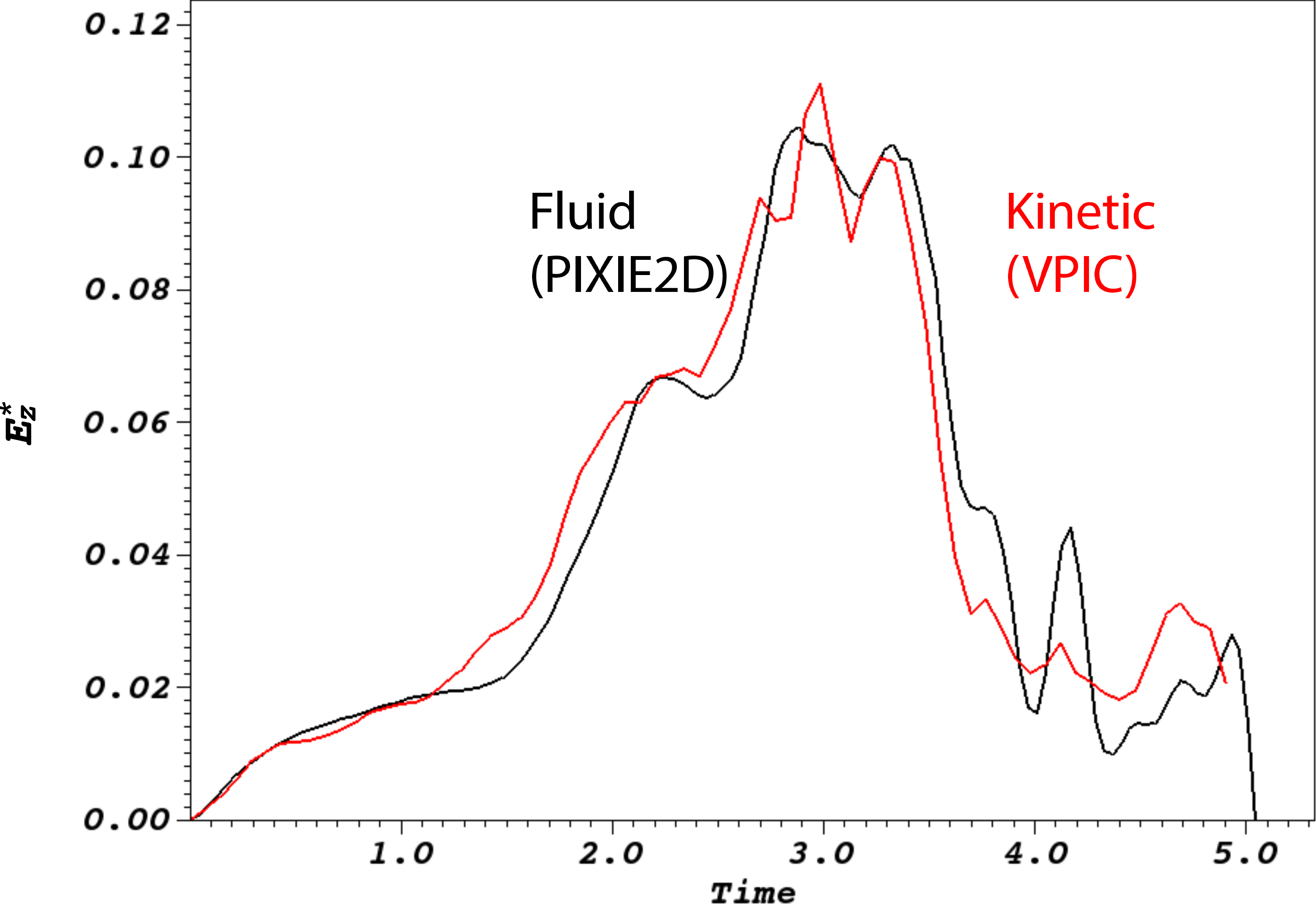}
\caption{\label{fig:rate-fluidvskinetic-de}Normalised reconnection rate $E_z^*$ against time for Harris-sheet run without FDWs ($d_e = 5\rho_s = 0.01$), from a fluid simulation (black) and kinetic simulation (red). The fluid simulation is the $\eta_H = 10^{-9}$ run from Fig.~\ref{fig:harris-dis}.}
\end{figure}

\begin{figure*}
\includegraphics[width=1.0\textwidth]{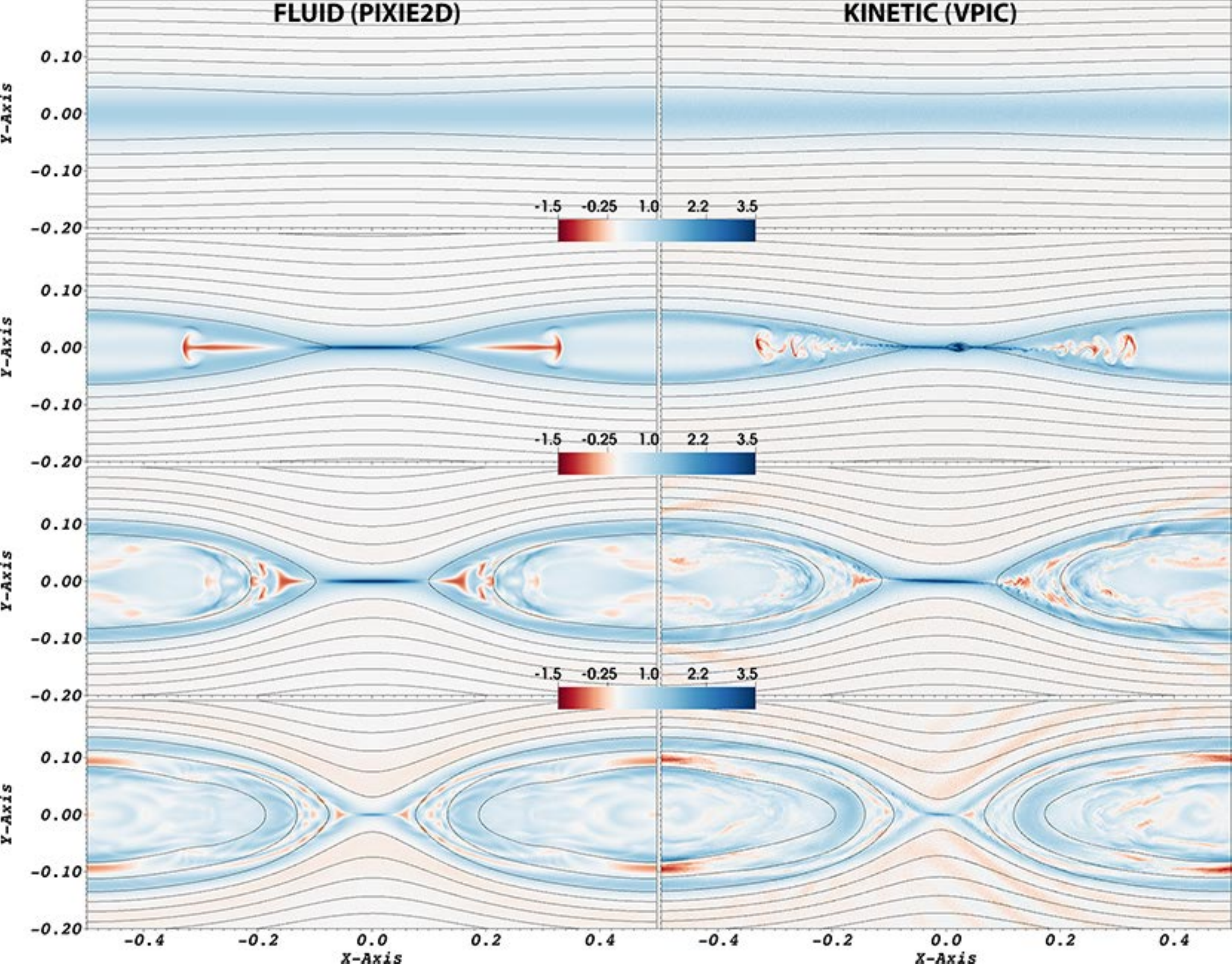}
\caption{\label{fig:snaps-fluidvskinetic-de}Snapshots of current density (colour) and magnetic flux (contours) from fluid (PIXIE2D) and kinetic (VPIC) simulations at $t=0$ (top), $t=1.6$, $t=2.64$ and $t=3.56$ (bottom). The set-up is the same as for the Harris-sheet reconnection simulations in Fig.~\ref{fig:harris-dis}, with $d_e = 5\rho_s = 0.01$. The current density is normalised by the value at the null-line at $t=0$.}
\end{figure*}

It has been demonstrated in cases with and without FDWs that the DR has the capacity to self-adjust, such that the reconnection rate becomes independent of DR physics. Therefore, reasonable agreement might be expected between fluid and kinetic rates within the regime of validity for Eqs.~(\ref{vorteqn},~\ref{vecBeqn}), namely with cold ions and strong guide-field. Fig.~\ref{fig:rate-fluidvskinetic-de} shows the reconnection rate from the $L_x = 0.5$ Harris-sheet simulation of Fig.~\ref{fig:harris-dis}, with $d_e = 5\rho_s = 0.01$ (no FDWs), $\mu = 10^{-5}$ and $\eta_H = 10^{-9}$, and a fully kinetic VPIC~\cite{bowers09} simulation using force-free initial conditions and with the same $d_e$ and $\rho_s$. Additional VPIC specific parameters used are a mass-ratio $m_i/m_e\,{=}\,15.2$, guide field to reconnecting field ratio $B_g\,{=}\,17.4$, electron thermal speed to speed-of-light ratio $v_{th,e}/c\,{=}\, 0.0625$, the ratio of electron plasma frequency to gyrofrequency using the upstream field is $\omega_{pe}/\Omega_{cex} \,{=}\, 10$, and the ratio of ion to electron temperature is $T_i/T_e\,{=}\,1/4$. There is remarkable agreement between the VPIC and fluid rate, which demonstrates that the essential physics is captured in Eqs.~(\ref{vorteqn},~\ref{vecBeqn}) and that the rate does not depend on the DR physics, which differs between the two descriptions. This latter point can be seen in Fig.~\ref{fig:snaps-fluidvskinetic-de}, which shows a comparison of the current density and flux evolution between these fluid and kinetic runs. The kinetic runs show different electron scale physics, such as a kinking of the electron outflow jet and formation of a small secondary island at $t=1.6$, which are not present in the fluid runs. However, at larger scales the two runs look very similar and, in particular, the evolution of the layer length is the same in both descriptions.  

\begin{figure}
\includegraphics[width=0.46\textwidth]{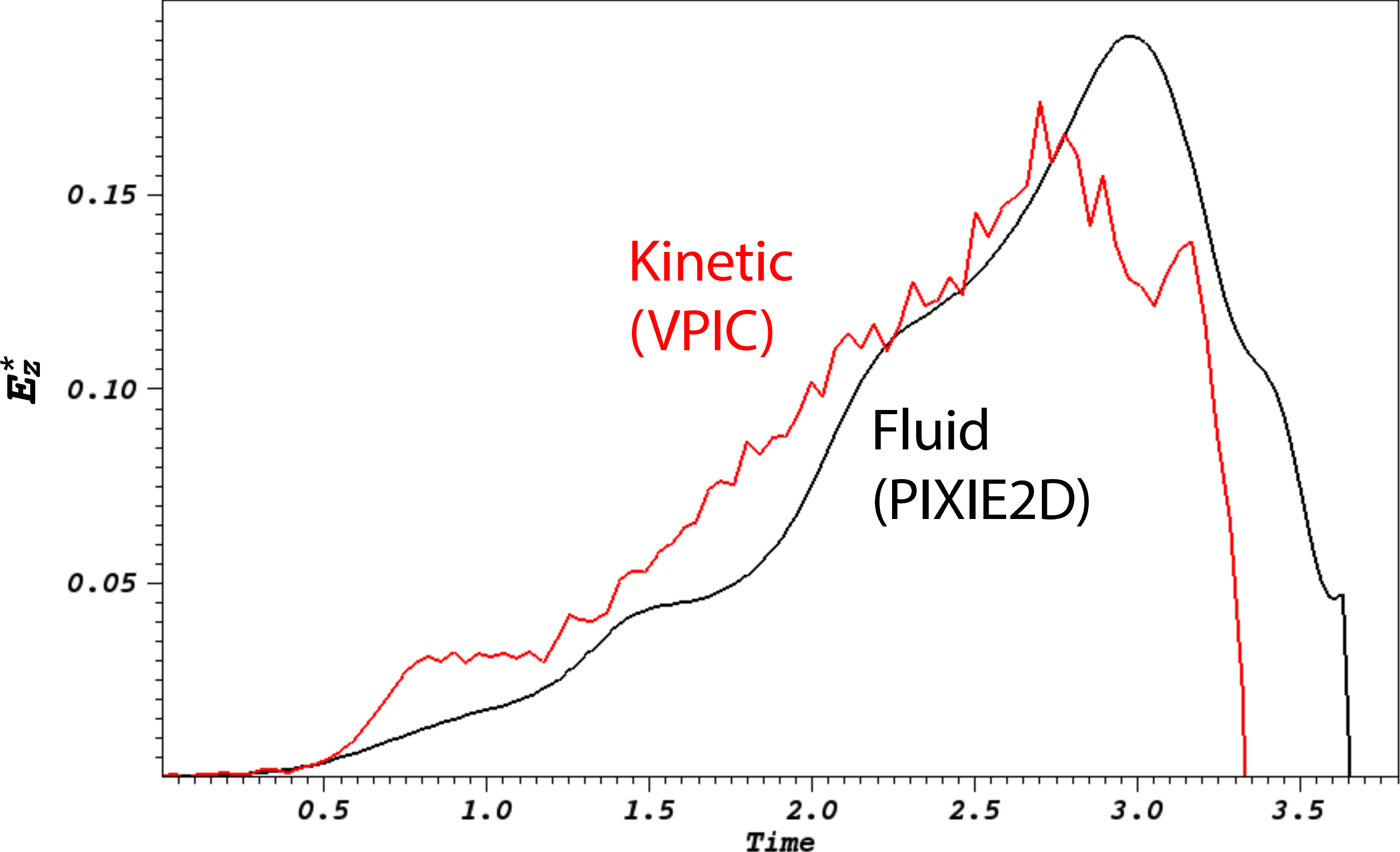}
\caption{\label{fig:fluidvskinetic-rate-rs}Normalised reconnection rate $E_z^*$ against time for Harris-sheet run with FDWs ($\rho_s = 5d_e = 0.01$), from a fluid simulation (black)  and kinetic simulation (red).}
\end{figure}

\begin{figure*}
\includegraphics[width=1.0\textwidth]{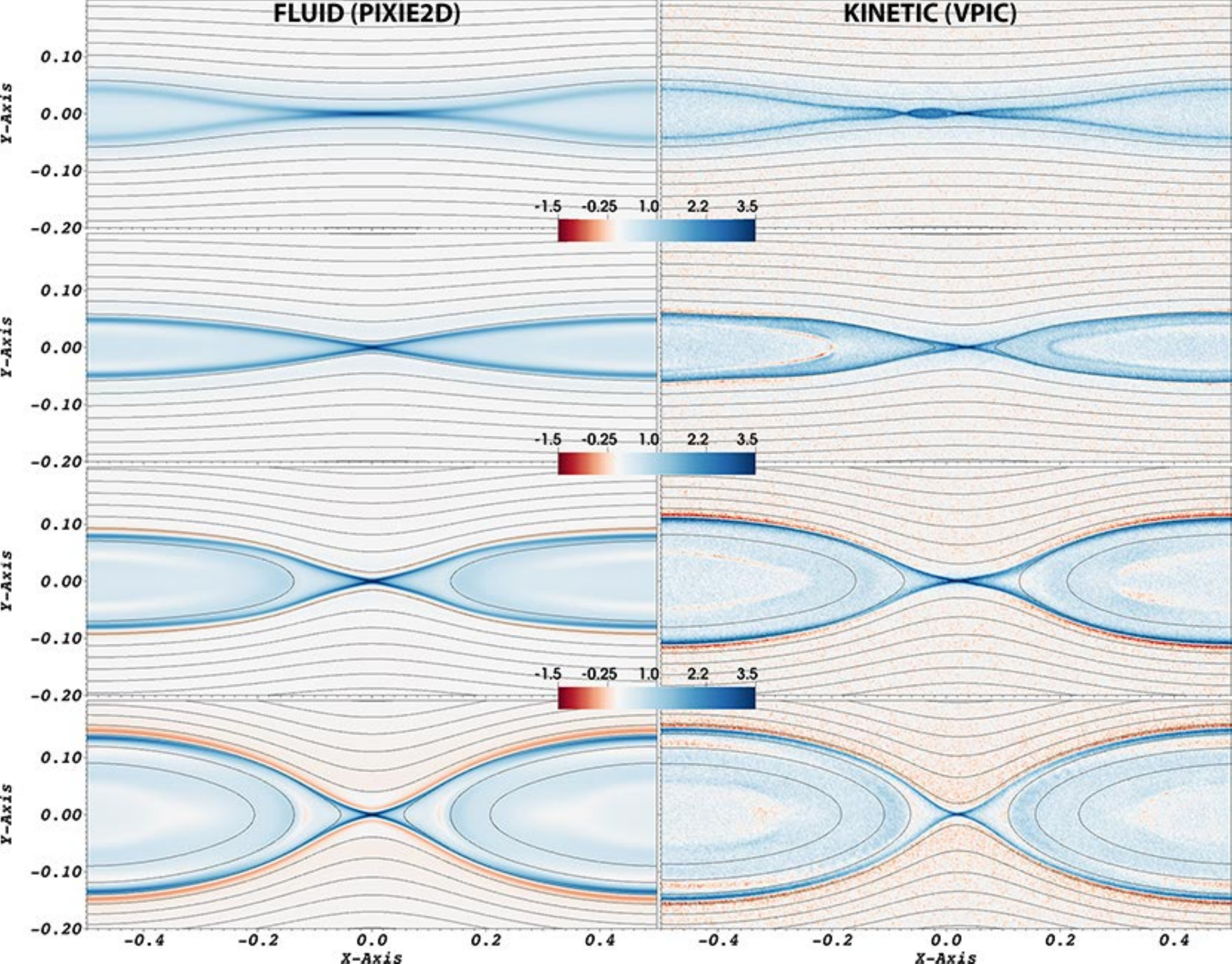}
\caption{\label{fig:snaps-fluidvskinetic-rs}Snapshots of current density (colour) and magnetic flux (contours) from fluid (PIXIE2D) and kinetic (VPIC) Harris-sheet simulations with $\rho_s = 5d_e = 0.01$ at $t=0.71$ (top), $t=1.4$, $t=2.11$ and $t=2.8$ (bottom). The current density is normalised by the value at the null-line at $t=0$.}
\end{figure*}

A number of studies have compared fluid and kinetic reconnection rates in the $\rho_s > d_e$ regime. Ref.~\onlinecite{rogers11} showed reasonably good agreement (within a factor of $\sqrt{2}$) in the non-linear rates between a cold-ion gyro-kinetic simulation and a (non-reduced) Hall-MHD formulation with electron inertia. Ref.~\onlinecite{ohia12} also found good agreement in the reconnected flux over time between a fully kinetic VPIC simulation and a Hall-MHD formulation with electron inertia. Ref.~\onlinecite{zacharias14} presented a comparison between gyrokinetic and reduced gyro-fluid simulations of the tearing instability. In the non-linear regime, excellent agreement was found between the two descriptions in the growth and saturation of the magnetic islands. Finally, Ref.~\onlinecite{loureiro13} found excellent agreement between reduced MHD and a novel fluid-kinetic model for the magnetic island flux at saturation in the large $\Delta '$ regime.

Figures~\ref{fig:fluidvskinetic-rate-rs} and \ref{fig:snaps-fluidvskinetic-rs} compare the reduced formulation in Eqs.~(\ref{vorteqn},~\ref{vecBeqn}) and fully kinetic VPIC simulations in the force-free Harris-sheet set-up of Fig.~\ref{fig:snaps-fluidvskinetic-de}, but with $\rho_s = 5d_e = 0.01$. Additional fluid parameters are $\eta_H = 10^{-9}$ and $\mu=10^{-4}$, and kinetic (VPIC) parameters are $m_i/m_e\,{=}\,380.3$, $B_g\,{=}\,12$, $v_{th,e}/c\,{=}\, 0.308$, $\omega_{pe}/\Omega_{cex} \,{=}\, 10$, and $T_i/T_e\,{=}\,1/10$. The plasma $\beta$ calculated with the in-plane magnetic field is large in this kinetic simulation, and so $16,000$ particles per cell were required to reduce the noise to an acceptable level. The normalised reconnection rate in both fluid and kinetic runs is $E_z^*$, the same as in Eq.~(\ref{harrisrate}), normalised by the Alfv\'enic rate at a distance of $4\rho_s$ upstream of the dominant X-point.

Fig.~\ref{fig:fluidvskinetic-rate-rs} shows at early time $(t \approx 0.6)$ that the kinetic rate has a slightly sharper increase than the fluid rate, and at late time the reconnection saturates slightly earlier in the kinetic simulation $(t=3.3)$ than the fluid simulation $(t=3.65)$. However, the overall agreement is very good, and in particular the peak rates are in agreement to within  $10\%$. 

The slight differences earlier on may be due to a secondary island that forms in the kinetic run at $t\approx0.5$ and is ejected at $t\approx0.7$. This island can be seen in Fig.~\ref{fig:snaps-fluidvskinetic-rs} at $t=0.71$. After this island ejection, the fluid and kinetic plots show good agreement in the structure of the current layer. Also, in both cases with and without FDWs, there is is reasonable agreement in the magnitude of the peak current density between fluid and kinetic runs. However, this quantity does depend on the dissipation, as we see larger peak current density in fluid runs with the lower hyper-resistivity $\eta_H = 10^{-11}$ (not shown).

It is also interesting to compare the structure of the current layer in Fig.~\ref{fig:snaps-fluidvskinetic-de}, where $d_e = 5\rho_s = 0.01$, with Fig.~\ref{fig:snaps-fluidvskinetic-rs}, where $\rho_s = 5d_e = 0.01$. In the former (without FDWs) the current layer is both thicker and longer than in the case with FDWs. Finally, we note that both the peak rates and the time at which reconnection saturates is within a factor of 2 between Fig.~\ref{fig:rate-fluidvskinetic-de} and Fig.~\ref{fig:fluidvskinetic-rate-rs}. 

\section{\label{sec:sys-size}System-size independence}

Historically, fast-reconnection has been defined as having a rate independent of the Lundquist number,~\cite{priest00} which corresponds to a rate independent of both collisional dissipation and system-size in the single-fluid regime. However, in the two-fluid regime, additional physical length scales are present and it is necessary to check the dissipation independence and the system-size independence separately. Here we present a system-size dependence study in the low-$\beta$ regime, considering the cases with FDWs $(\rho_s^2 \gg d_e^2)$ and without FDWs $(d_e^2 \gg \rho_s^2)$. For this study, we use the Harris-sheet reconnection problem set-up, as it is not yet clear whether there is good agreement between fluid and kinetic simulations of island coalescence, particularly for large islands, see e.g. Ref.~\onlinecite{karimabadi11}. 

The requirements for a system-size independent reconnection rate can be understood by considering how the local two-fluid region parameters in Eq.~(\ref{bothscalings}) vary with system-size,  assuming that both the DR and outer two-fluid regions are approximately in steady-state at the time the rate is measured. For the study presented in this section, we vary the length of the equilibrium current layer, $L_x$, and keep the box size in the inflow direction, $L_y$, and the ratio of the two-fluid scale to equilibrium current sheet thickness, $h/\lambda$, fixed. With this, the upstream field $B_{xh}$ does not vary appreciably with $L_x$, and so the condition for system-size independence becomes $w_v \propto L_x^0$. If there is no mechanism to limit $w_v$, then Eq.~(\ref{bothscalings}) suggests that the rate will decrease with system-size as $E_z \propto w_v^{-1} \propto L_x^{-1}$.

\begin{figure}
\includegraphics[width=0.45\textwidth]{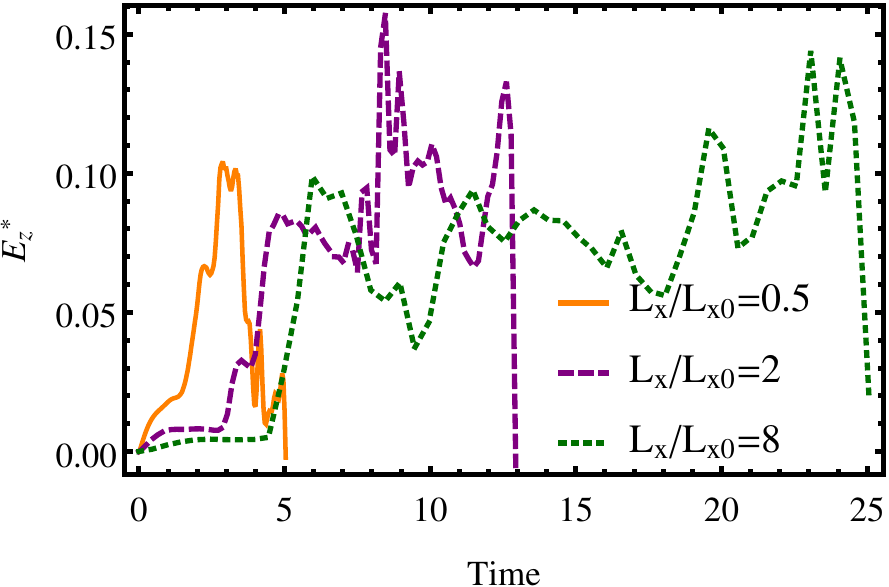}
\caption{\label{fig:sys-size-derates}Normalised reconnection rates $E_z^*$ against time for fluid Harris-sheet simulations with $d_e =5\rho_s =0.01$, $\mu =10^{-4}$ and $\eta_H =10^{-9}$. Shown are $L_x/L_{x0} = 0.5$ (orange solid, same as in Fig.~\ref{fig:rate-fluidvskinetic-de}), $L_x/L_{x0}=2$ (purple, dashed), and $L_x/L_{x0} = 8$ (green, dotted) where $L_{x0} = 100 d_e$.}
\end{figure}

Figure~\ref{fig:sys-size-derates} shows the normalised reconnection rate $E_z^*$, see Eq.~(\ref{harrisrate}), against time from Harris-sheet fluid simulations with the same set-up as in Figs.~\ref{fig:harris-dis}, \ref{fig:rate-fluidvskinetic-de}, but with varying $L_x$. In contrast to the $L_x = 0.5$ run, where there is no secondary island formation, the sharp increase in the rate for the $L_x = 2$ and $L_x = 8$ runs occurs along with the formation of secondary islands, at $t\approx3.2$ and $t\approx5$ respectively. At later times there is significant secondary island formation in both of these runs (see below), and the rates are somewhat spiky. Although the runs with larger $L_x$ take longer to reach the initial sharp increase in the rate, there is no clear dependence of the rate on the system-size $L_x$ after this onset. 

\begin{figure}
\includegraphics[width=0.45\textwidth]{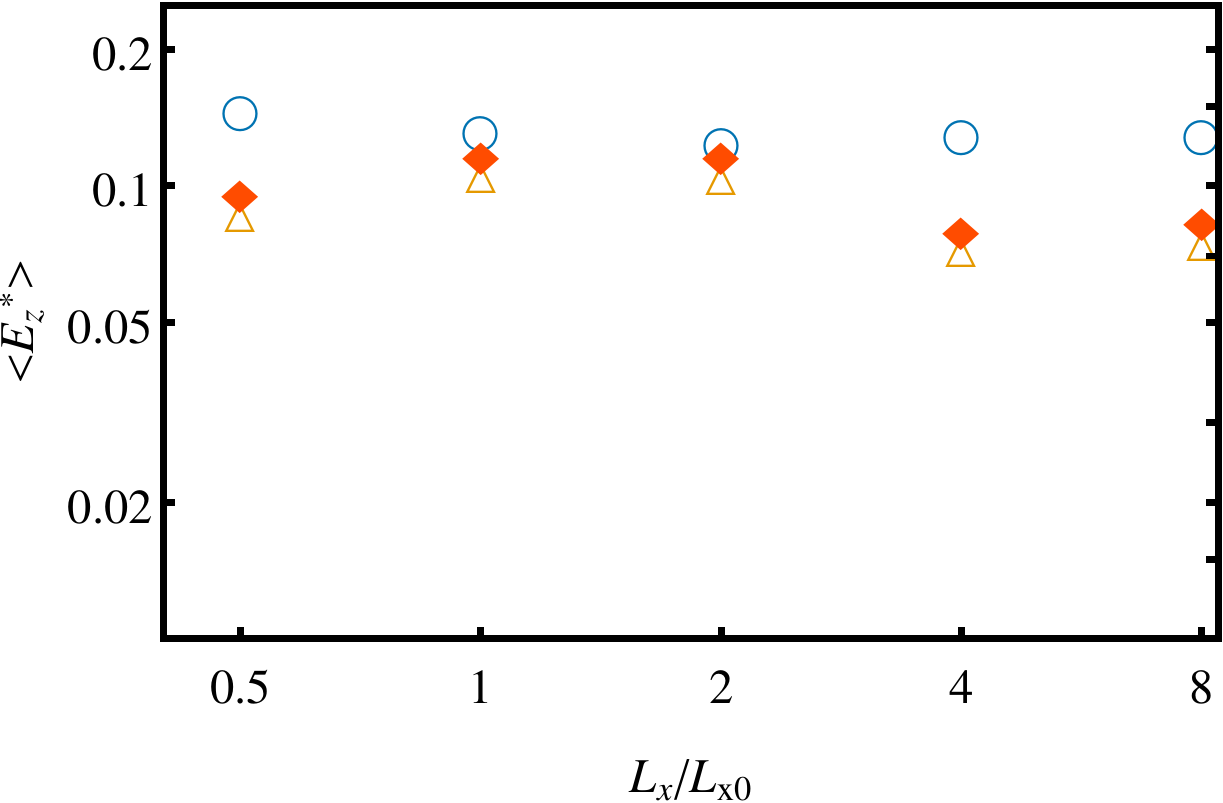}
\caption{\label{fig:sys-size-avpeak}Averaged peak rate $<E_z^*>$ against system-size $L_x/L_{x0}$, where $L_{x0} = 100 h$. From Harris-sheet fluid simulations with $\mu=10^{-5}$, $\eta_H = 10^{-9}$, $d_e = 5\rho_s = 0.01$ (orange $\triangle$), $\rho_s = 5 d_e = 0.01$ (blue $\bigcirc$), and fully kinetic simulations with $d_e = 5\rho_s = 0.01$ (red $\blacklozenge$). Reprinted with permission from Stanier et al., Phys. Plasmas, 22, 010701, (2015). Copyright (2015) American Institute of Physics.}
\end{figure}

Figure~\ref{fig:sys-size-avpeak} shows the peak reconnection rates $<E_z^*>$ plotted against system-size, with $E_z^*$ as in Eq.~(\ref{harrisrate}) and normalised by the Alfv\'enic rate at $4h$ upstream of the dominant X-point (this distance is chosen to be outside of the initial field-reversal region of thickness $\lambda$). Here, we average the rate ($<>$) over the time required for an Alfv\'en wave to cross the whole box in the inflow direction, so that the measurement of the peak rate is not overly affected by sharp spikes in the rate associated with secondary island formation, see Fig.~\ref{fig:sys-size-derates}. Also, we neglect the final $25 \%$ of the simulation time when calculating this averaged peak rate as the upstream magnetic field can become significantly depleted close to saturation, leading to large increases in the normalised rate. 

In Fig~\ref{fig:sys-size-avpeak}, we show results from: fluid simulations with $\rho_s = 0.01$, $d_e = 0.002$ (with FDWs); fluid runs with $d_e = 0.01$, $\rho_s = 0.002$ (without FDWs); and VPIC simulations with $d_e = 0.01$ and $\rho_s = 0.002$. The fluid simulations all have $\mu = 10^{-5}, \,\eta_H = 10^{-9}$, while the additional VPIC specific parameters are the same as in Sec.~\ref{sec:comparekinetic}, but with $\omega_{pe}/\Omega_{cex} = 2.5$ and $v_{th,e}/c = 0.25$. This change in the VPIC specific parameters gives runs that are less expensive than the one in Sec.~\ref{sec:comparekinetic}, particularly for the large simulations presented here, but the Alfv\'en speed becomes mildly relativistic. Thus, we normalise by the relativistic Alfv\'enic rate for the VPIC runs, see Ref.~\onlinecite{liu14}. 

Firstly, the case with FDWs has a clear system-size independent rate, with $<E_z^*> \approx 0.15$. For the case without FDWs, there is some small variation in the peak rate that appears due to differences in secondary island formation between runs, even after the time averaging of the rate. However, there is no appreciable downward trend as the rate for the $L_x = 0.5$ (total box length of $100 d_e$) simulation is very close to the $L_x = 8$ ($1600 d_e$) run. As the rate is a significant fraction of the Alfv\'enic rate for cases with and without FDWs, both cases satisfy all the criteria of fast reconnection; they are Alfv\'enic, independent of dissipation, and of system-size. In addition, there is good agreement in the peak rates between VPIC and fluid calculations for the runs without FDWs. However, we note that for larger $L_x$ the time histories of $E_z^*$ do not agree quite as well as in Fig.~\ref{fig:rate-fluidvskinetic-de}, due to the sensitivity of secondary island formation on factors such as the level of noise (not shown).

\begin{figure*}
\includegraphics[width=1.0\textwidth]{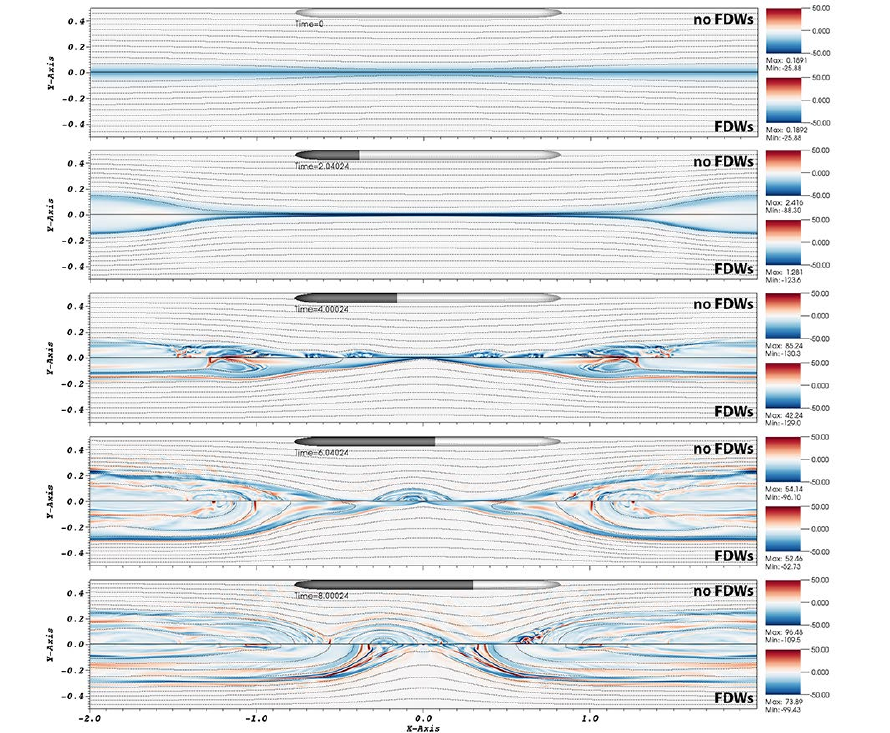}
\caption{\label{fig:lx2-currentflux}Current and flux for $L_x=2$ from PIXIE2D fluid simulations. Top half of each snapshot is for $d_e = 0.01$, $\rho_s = 0.002$ (no FDWs), and bottom half is $\rho_s = 0.01$, $d_e = 0.002$ (FDWs).}
\end{figure*}

Previous low-$\beta$ kinetic simulations have found that the absolute values for the reconnection rates in simulations without FDWs can be as large as,~\cite{liu14} or even exceed,~\cite{tenbarge14} those for simulations with FDWs. In this set of Harris-sheet simulations we find that the rate is typically larger for the simulations with $\rho_s = 5d_e = 0.01$ (with FDWs) than those with $d_e = 5\rho_s = 0.01$ (without FDWs), although this difference is within a factor of two. We also note that for the island coalescence runs in Fig.~\ref{fig:ic-rates}, and between the Harris sheet runs with $\lambda=0.5$ in Figs.~\ref{fig:rog-rates}a and~\ref{fig:rog-rates}b of Sec.~\ref{sec:rogerscomparison},  the rates are much closer between the two cases. 

\begin{figure*}
\includegraphics[width=1.0\textwidth]{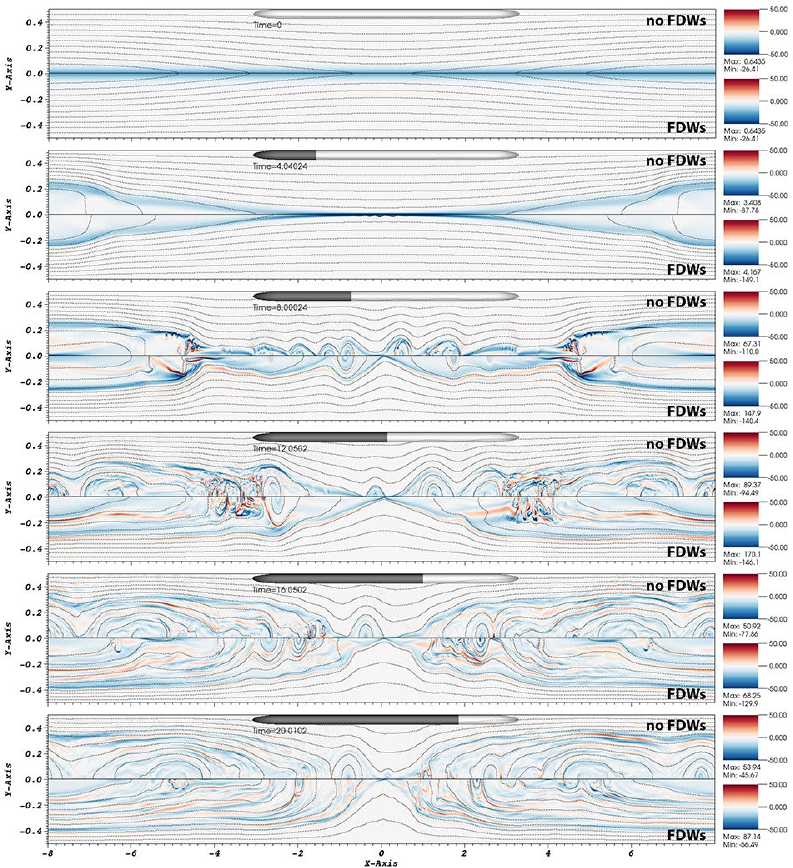}
\caption{\label{fig:lx8-currentflux}Current and flux for $L_x=8$ from PIXIE2D fluid simulations. Top half of each snapshot is for $d_e = 0.01$, $\rho_s = 0.002$ (no FDWs), and bottom half is $\rho_s = 0.01$, $d_e = 0.002$ (FDWs).}
\end{figure*}

Figures~\ref{fig:lx2-currentflux} and~\ref{fig:lx8-currentflux} show several time snapshots of the current density and magnetic flux from PIXIE2D fluid simulations in cases without (top halves of each time snapshot) and with (bottom halves of each time snapshot) FDWs, for the $L_x = 2$ and $L_x = 8$ runs respectively. For $L_x = 2$ there are clear qualitative differences at $t\ge 4$, after the initial perturbation has compressed the current layer. For the case without FDWs the current layer shows multiple islands at $t=4$, and a large island at $t=6, 8$, which appears to limit $w_v$. In contrast to this, the current layer in the case with FDWs does not produce a secondary island until $t\approx 8$, where the reconnection is close to saturating. The structure of the flux and current in the case with FDWs does resemble the open X-point  configuration~\cite{petschek64} more closely than the case without FDWs in this set of runs. Such an open X-point configuration has been linked to~\cite{grasso99, comisso12} the conservation of the generalised fields $G_{\pm}  = \psi - d_e^2 j \pm d_e \rho_s \omega$, which in the collisionless limit ($\eta_H = \mu=0$) are Lagrangian conserved quanities in the flow fields $\boldsymbol{v}'_{\pm} = \boldsymbol{\hat{z}}\times\boldsymbol{\nabla} \phi'_{\pm}$, with $\phi'_{\pm} = \phi \pm \rho_s \psi/d_e$. Here we have finite dissipation so that these generalised fields can reconnect, but we do see that $G_{\pm}$ is advected by the counter-twisted flows $\boldsymbol{v}'_{\pm}$ to give localised current and vorticity structures along the separatrices of $\phi'_{\pm}$, as in Ref.~\onlinecite{grasso99}. Interestingly, Figure 5 of Ref.~\onlinecite{grasso99} shows that the X-point begins to open up even when $\rho_s = d_e$ and there are no FDWs present. The opening angle between the separatrices has been shown to increase in the presence of hot ions.~\cite{comisso13}

For the $L_x = 8$ runs, the differences are less clear, as there are multiple islands at $t=8$ in cases with and without FDWs. However, the opening angle between the separatrices at $t=8$ does appear larger in the case with FDWs, whereas there is more secondary island formation at late times ($t>8$) in the case without FDWs. Finally, despite these qualitative differences, the time at which reconnection saturates is very similar between cases with ($t \approx 20$) and without ($t \approx 25$) FDWs, at which point a similar amount of flux has been reconnected.

It is not clear whether the mechanism for limiting $w_v$ to give system-size independent reconnection rates differs between the cases with and without FDWs. There are significant qualitative differences in structure of the current layer between these cases in Figs.~\ref{fig:lx2-currentflux},~\ref{fig:lx8-currentflux}, where it appears that secondary island formation plays a role in regulating the length of the layer in the case without FDWs. However, the reconnection rates in these $L_x = 2, 8$ runs without FDWs are within a factor of two of the rates for the runs with FDWs, and also similar to the $L_x = 0.5$ fluid simulation without FDWs, where there is no secondary island formation. 

\section{\label{sec:rogerscomparison}Comparison with previous two-fluid study}
In this section, the results from a set of two-fluid and kinetic simulations are presented that can be more closely compared with earlier two-fluid results from Fig.~3a of Ref.~\onlinecite{rogers01}. Here, fluid simulations are performed in a quarter-box $(x,y) \in [0,10]\times [0,5]$ using the Harris-sheet initial conditions of Eq.~(\ref{initialcondha}). Two different values for the equilibrium current sheet thickness are used,  $\lambda = 0.5$ and $\lambda = 1$.

Figure 3a of Ref.~\onlinecite{rogers01} shows the reconnection rate against $\beta_x/2$ (the plasma-$\beta$ defined in terms of the upstream reconnecting field $B_{x0}$) for a fixed guide to reconnecting field ratio of $B_0/B_{x0} = 30$, and a fixed mass-ratio of $m_i/m_e = 82.6$. Here, we match these parameters in our reduced model by choosing $d_e = 1/\sqrt{m_i/m_e} = 0.11$, and vary $\rho_s$, since $\rho_s \propto (\beta_x/2)^{1/2}$ with $B_0/B_{x0}$ and $m_i/m_e$ fixed. The threshold for the plasma to support fast-dispersive kinetic Alfv\'en waves is $\rho_s > d_e$ as before, which corresponds to $\beta_x/2 > (m_e/m_i) (B_0^2/B_{x0}^2) = 10.9$, in agreement with the threshold (dotted line) given in Fig. 3a of Ref.~\onlinecite{rogers01}. For the PIXIE2D fluid simulations, we use $\eta_H = 10^{-6}, \mu = 10^{-4}$, which ensure that the DR thickness is sufficiently below the two-fluid scales.

\begin{figure}
\includegraphics[width=0.45\textwidth]{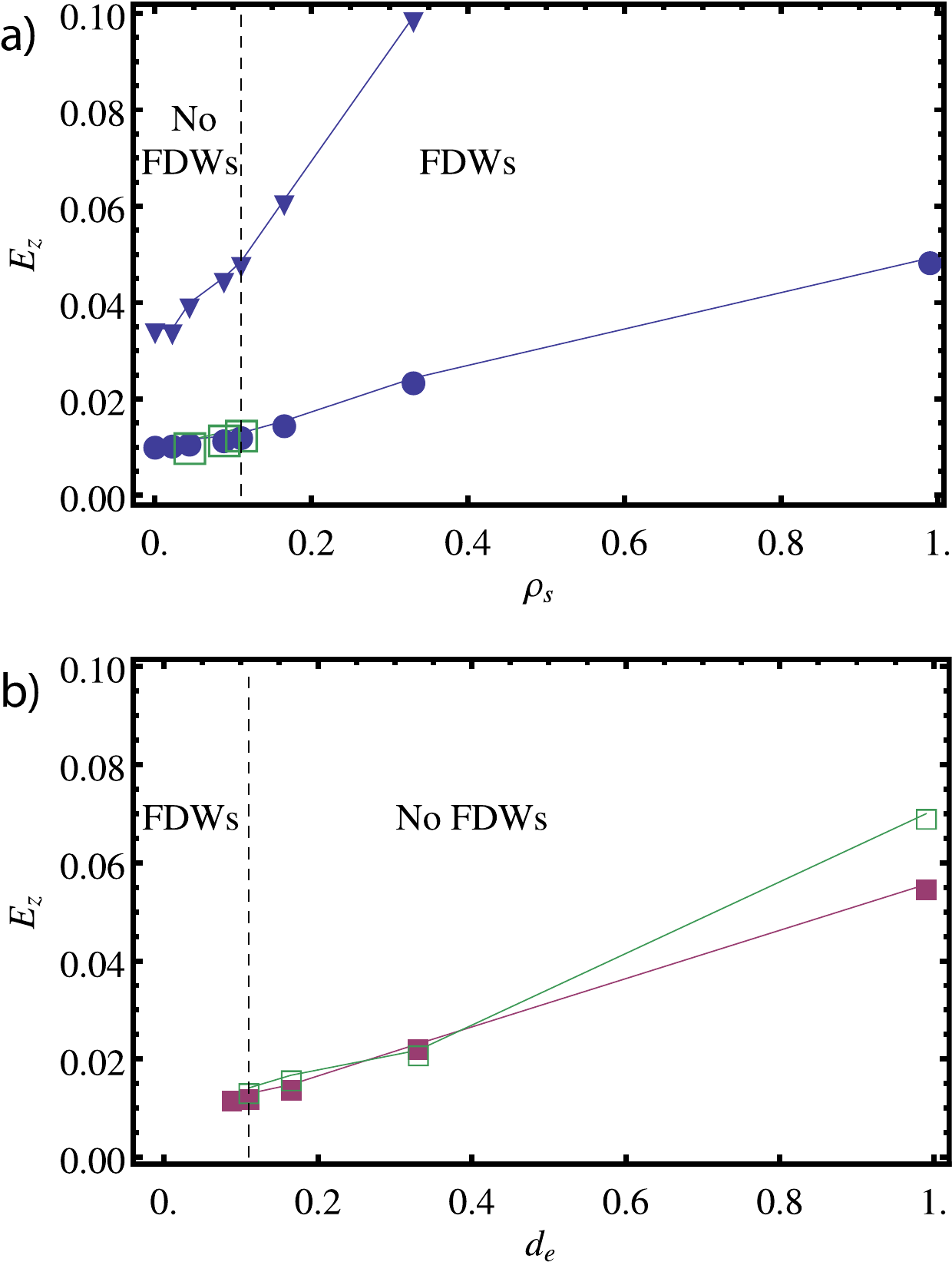}
\caption{\label{fig:rog-rates}a) Reconnection rate $E_z$ from Harris-sheet simulations with fixed $d_e = 0.11$ and different $\rho_s$. Shown are fluid simulations with initial current sheet thickness $\lambda = 0.5$ (blue $\bigtriangledown$), and fluid (blue $\bigcirc$) and kinetic (green $\Box$) simulations with $\lambda = 1$. b) $E_z$ for fixed $\rho_s = 0.11$ and varying $d_e$ for fluid (magenta filled $\Box$) and kinetic (green hollow $\Box$) simulations with $\lambda = 1$.}
\end{figure}

Fig.~\ref{fig:rog-rates}a shows the reconnection rate $E_z$, measured at the dominant X-point and normalised by the Alfv\'enic rate at the upstream boundary $B_{x0}^2$, plotted against $\rho_s$. Here, blue $\bigtriangledown$ are from PIXIE2D fluid simulations with initial current sheet thickness $\lambda = 0.5$ (the thickness used in Ref.~\onlinecite{rogers01}), blue $\bigcirc$ are from PIXIE2D fluid simulations with $\lambda = 1$, and green $\Box$ are from VPIC simulations with $\lambda = d_i = 1$, $B_0/B_{x0} = 30$, $m_i/m_e = 82.6$ and varying $\rho_s$. As was found in Ref.~\onlinecite{rogers01}, we see the reconnection rate decrease with $\rho_s\propto (\beta_x/2)^{1/2}$ such that the rate is lower in the region where there are no FDWs than with FDWs. However, there are some subtle differences between the $\lambda = 0.5$ runs presented here compared with Ref.~\onlinecite{rogers01} that are worth highlighting.

Firstly, it is mentioned in Ref.~\onlinecite{rogers01} that there is a factor of two decrease in the rate between a regime with both whistler and  kinetic Alfv\'en waves and a regime with kinetic Alfv\'en waves only. In Fig. 3a of Ref.~\onlinecite{rogers01}, the maximum $\beta_x/2 = 30$ which corresponds to $\rho_s \approx 0.18$ in our Fig.~\ref{fig:rog-rates} a. We find that for the simulation with larger $\rho_s = 0.33$, such that $\rho_s/\lambda$ is closer to unity, the rate $E_z \approx 0.1$ even without whistler waves. 

Secondly, we find that the decrease in the rate with $\rho_s$ in the region $\rho_s < d_e$ (No FDWs) is slightly flatter than that found in Ref.~\onlinecite{rogers01}, such that for $\rho_s = \beta_x = 0$ we find $E_z \approx 0.033$, which is more than twice the value reported in Ref.~\onlinecite{rogers01}. In these simulations (with $\lambda = 0.5$) we do see some secondary islands for the $\rho_s = 0, 0.022, 0.044$ runs that may effect the rates in this region. Note that for these runs we repeat the simulation in the half-box $(x,y) \in [-10,10]\times [0,5]$ to prevent islands from being trapped at $x=0$.  

We do not see any secondary island formation in any of the simulations reported here with the thicker initial current sheets $\lambda =1$ (blue $\bigcirc$ in Fig.~\ref{fig:rog-rates} a). The peak rates for these runs are smaller in magnitude, but we see the same overall trend with a flattening of the rate when $\rho_s < d_e$, which is also shown in fully kinetic VPIC simulations (green $\Box$), and there is a sharper increase in the rate when $\rho_s > d_e$. This trend is consistent with the idea that the reconnection rate depends only on the maximum of the two-fluid scales $h=\textrm{max}[\rho_s,d_e]$.

Figure~\ref{fig:rog-rates} b shows the reconnection rate from a set of Harris sheet simulations with $\lambda =1$ using the same set-up, but with fixed $\rho_s = 0.11$ (dashed line) and varying $d_e$. For these simulations, kinetic Alfv\'en waves are supported to the left of the dashed line, while there are no FDWs to the right of the dashed line.  Remarkably, the peak rates increase with increasing $d_e$, with almost the same values as for increasing $\rho_s$ with fixed $d_e$ in Fig.~\ref{fig:rog-rates} a, supporting the conclusion that the rate depends only on the maximum of the two-fluid scales, regardless of the presence of FDWs. Fully kinetic (VPIC) runs with the same fixed $\rho_s$ and varying $d_e$ also show the same trend (green $\Box$). 

\begin{figure}
\includegraphics[width=0.5\textwidth]{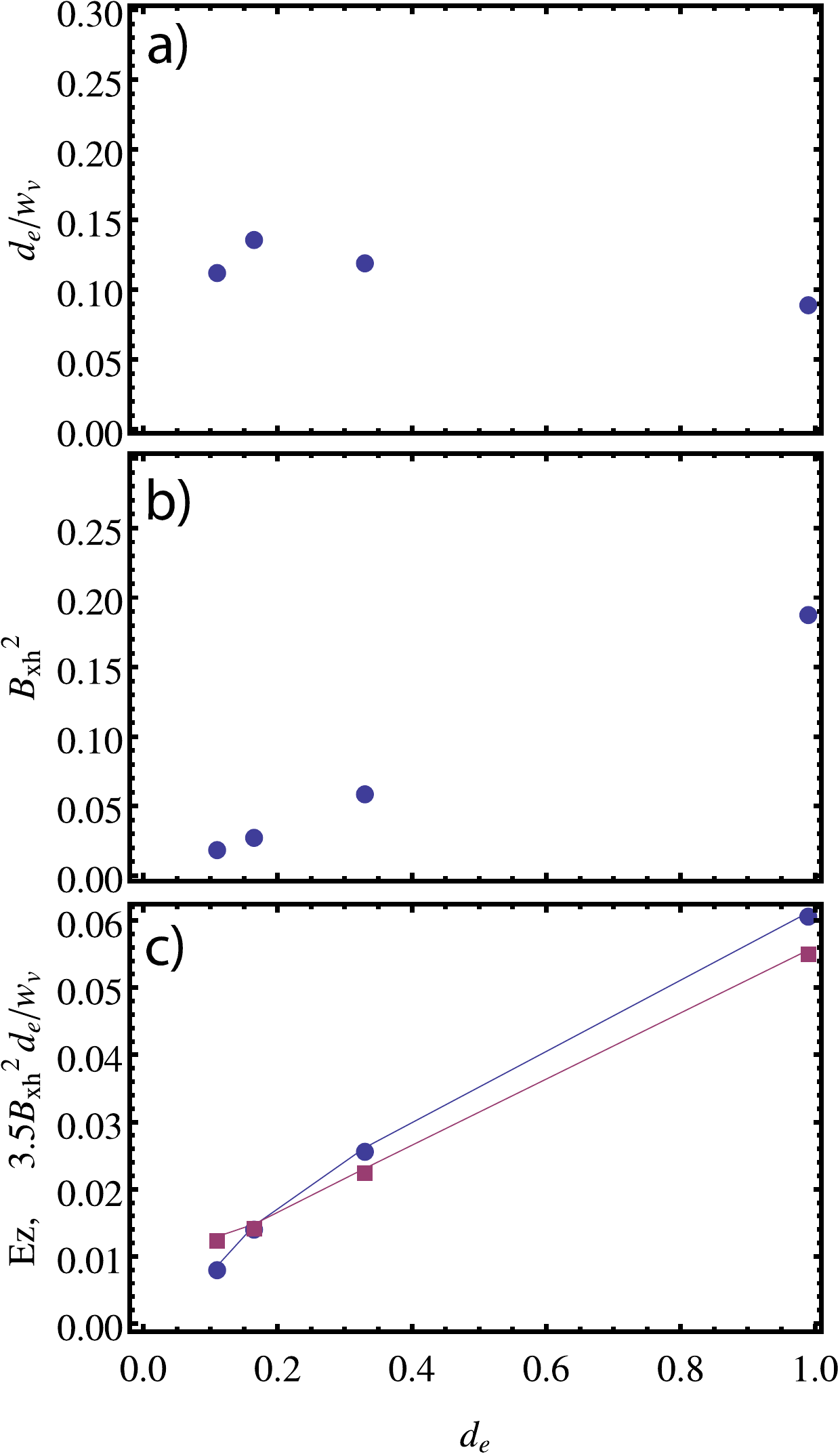}
\caption{\label{fig:rogtfparams}a) Aspect ratio of outer two-fluid region $d_e/w_v$, b) Alfv\'enic rate, $B_{xh}^2$, on edge of two-fluid region, c) Actual rate at X-point $E_z$ (magenta $\Box$), and rate from Eq.~(\ref{bothscalings}) multiplied by $3.5$ (blue $\bigcirc$). All results from $\lambda = 1$ fluid simulations with fixed $\rho_s = 0.11$.}
\end{figure}

To understand why the rate increases for fixed $\rho_s$ and increasing $d_e$, Fig.~\ref{fig:rogtfparams} shows the two-fluid region parameters that govern the rate in Eq.~(\ref{descalings}). Interestingly,  Fig.~\ref{fig:rogtfparams}a shows the aspect-ratio of the outer region $d_e/w_v$, at the time of peak rate, remains roughly constant with value $\approx 0.1$ over a large range in $d_e$. However, the Alfv\'enic rate on the edge of the two-fluid region at the time of peak rate, $B_{xh}^2$, in Fig.~\ref{fig:rogtfparams}b does increase with $d_e$. This accounts for the scaling of the reconnection rate, as shown in~\ref{fig:rogtfparams}c, where the expression for the rate in Eq.~(\ref{descalings}) has been scaled by a numerical factor of $3.5$.

We find that for runs with a smaller ratio of the maximum two-fluid scale to equilibrium current sheet thickness, $h/\lambda$, there is a longer onset period before the peak reconnection rate is reached, and the upstream Alfv\'enic rate $B_{xh}^2$ has time to significantly decrease. This is very similar to the results found in a previous study of Hall-MHD reconnection with large equilibrium current sheet thickness.~\cite{shay04} In both cases, with and without FDWs, if the peak rate is normalised by the instantaneous value of $B_{xh}^2$, then it remains roughly constant against $h$. From the results presented in this section, we argue that the presence or absence of FDWs does not play a primary role in whether the magnitude of the reconnection rate is large, as was concluded previously in Ref.~\onlinecite{rogers01}.

\section{\label{sec:conclude}Summary and conclusions}

In this paper we show that low-$\beta$ two-fluid reconnection is formally fast, regardless of the presence of fast-dispersive waves (FDWs) that have previously been suggested to play a critical role.~\cite{rogers01} This paper gives additional details on the method used, and gives additional evidence to support the conclusions in a recently published letter.~\cite{stanier15} We show that reconnection in Harris sheet geometry is independent of the physics of the dissipation region even in the absence of FDWs, and demonstrate for cases with and without FDWs that the reconnection rate and the overall length of the current layer are in very good agreement between fluid and kinetic simulations.

For simulations with very large system-sizes, it is not yet clear which mechanism limits the length of the layer $w_v$ and, in particular, whether the mechanism is the same regardless of the presence of FDWs. In Sec.~\ref{sec:sys-size} we do see qualitative differences between the structure of the layer, such as a larger opening angle between the separatrices in the case with FDWs, and more secondary islands in the case without FDWs. However, the peak rates between the two cases agree to within a factor of two, and the time until saturation is also very similar. Furthermore, in the case without FDWs, the reconnection rates for small systems ($L_x = 0.5$) with no secondary islands are the same as that for the larger systems ($L_x = 2, 8$) which do have secondary islands. Interestingly, for the Harris-sheet simulations without FDWs presented in Fig.~\ref{fig:rogtfparams}, there is no secondary island formation and the aspect ratio of the layer is found to be $d_e/w_v \approx 0.1$ across a large range in $d_e$. This suggests that there is a mechanism other than secondary island formation that can limit the length of the layer $w_v$ in the case without FDWs. 

Also in Sec.~\ref{sec:rogerscomparison}, we argue that the decrease in the reconnection rate with $\beta_x/2$ is not related to the presence of fast-dispersive waves, which was concluded in a previously two-fluid study.~\cite{rogers01} Instead, we argue that it is due to the decrease in the upstream Alfv\'enic rate $B_{xh}^2$ associated with a longer onset period for runs with smaller values of $h/\lambda$. This phenomenon is similar to that found in a previous Hall-MHD study, which considered how the reconnection rate scaled for small $d_i/\lambda$.~\cite{shay04}

\begin{acknowledgments}
This work is supported by the U.S. Department of Energy, Office of Science, Office of Fusion Energy Sciences, and used resources provided by the Los Alamos National Laboratory Institutional Computing Program, which is supported by the U.S. Department of Energy National Nuclear Security Administration under Contract No. DE-AC52-06NA25396. A.S. would like to thank Yi-Hsin Liu for helpful discussions, and the anonymous referee for their constructive suggestions that helped to improve this paper.
\end{acknowledgments}

%\nocite{*}
%\bibliography{pop-draft1}% Produces the bibliography via BibTeX.


\end{document}